\documentclass[twocolumn,trackchanges]{aastex701}

\def\hal{H$\alpha$}
\def\hb{H$\beta$}

\def\oiii{[\ion{O}{3}]}

\def\nii{[\ion{N}{2}]}

\def\h{\hspace{-2.0 mm}}
\def\asec{$^{\prime\prime}$}

\begin{document}

\title{Mid-infrared Variability-based AGN Selection using the Multi-epoch Photometric Data from WISE}

\author[0009-0009-9529-514X]{Shinyu Kim}
\affiliation{Department of Astronomy, Yonsei University, 
50 Yonsei-ro, Seodaemun-gu, Seoul 03722, Republic of Korea}
\affiliation{Department of Astronomy and Atmospheric Sciences, 
Kyungpook National University, Daegu 41566, Republic of Korea}
\email[]{}

\author[0000-0002-3560-0781]{Minjin Kim}
\affiliation{Department of Astronomy, Yonsei University, 
50 Yonsei-ro, Seodaemun-gu, Seoul 03722, Republic of Korea}
\email[]{mkim.astro@yonsei.ac.kr}

\author[0000-0002-5346-0567]{Suyeon Son}
\affiliation{Kavli Institute for Astronomy and Astrophysics, 
Peking University, Beijing 100871, People's Republic of China}
\email[]{}

\author[0000-0001-6947-5846]{Luis C. Ho}
\affiliation{Kavli Institute for Astronomy and Astrophysics, Peking University, Beijing 100871, People's Republic of China}
\affiliation{Department of Astronomy, School of Physics, Peking University, Beijing 100871, People's Republic of China}
\email{lho.pku@gmail.com}

\correspondingauthor{Minjin Kim}
\email{mkim.astro@yonsei.ac.kr}

\begin{abstract}
We assess the systematics and efficiency of an AGN selection method based on mid-infrared (MIR) variability. To this end, we utilize various types of active and inactive galaxies from the Sloan Digital Sky Survey, matching them with multi-epoch photometric data from the NEOWISE mission. Using W1 and W2 band light curves with a $\sim10$-year baseline, we find that combining the likelihood of deviation from non-variability with the correlation coefficient between the W1 and W2 bands reliably identifies AGNs. Specifically, this MIR-based method recovers $\sim 28.2\%$ of optically selected AGNs. Applying the same technique to inactive galaxies, we identify AGN candidates at fractions ranging from $0.4$ to $11.8\%$, indicating that MIR variability allows us to detect AGN candidates even in optically inactive hosts. While some variable sources exhibit transient-like light curves, possibly originating from tidal disruption events or supernovae, their contribution to the total variable population is less than a few percent, indicating a minimal impact on our results. Across all subsamples, the AGN fraction marginally increases with star formation activity, implying coordinated evolution between central black hole growth and star formation. Finally, the AGN fraction inferred from our method drops dramatically in classical LINERs, consistent with their low accretion rates and absence of a dusty torus.

\end{abstract}


\section{INTRODUCTION} 

The demography of active galactic nuclei (AGNs) is of great importance for understanding not only the formation and evolution of supermassive black holes (SMBHs) at the centers of massive galaxies \cite[e.g.,][]{ho_1995,ho_1997}, but also the evolution of the galaxies themselves, likely through AGN-driven feedback on star formation \cite[e.g.,][]{silk_1998, kormendy_2013}. Therefore, identifying AGN activity within large galaxy samples is a crucial component of investigating the co-evolution of galaxies and SMBHs. To this end, previous studies have developed various methods to detect AGNs using survey datasets. Modern spectroscopic surveys, such as the Sloan Digital Sky Survey (SDSS; \citealt{sdssdr1_2003}), enable the classification of AGNs based on their optical spectral properties. For example, a strong UV continuum and broad emission lines are characteristic of Type 1 AGNs \cite[e.g.][]{richards_2002}, whereas obscured (Type 2) AGNs are primarily identified via line ratios between Balmer emissions and various forbidden lines, reflecting the harder and more intensive ionizing continuum of AGNs relative to inactive galaxies \cite[e.g.,][]{baldwin_1981, veilleux_1987,kewley_2001}.

Although AGN selection methods based on optical spectroscopy are widely used, they are biased against low-luminosity and highly obscured AGNs \cite[e.g.,][]{imanishi_2009, agostino_2019}. Because highly obscured AGNs are key to probing the AGN population, particularly in the early Universe, alternative methods are required \cite[e.g.,][]{treister_2006}. One particularly effective approach utilizes mid-infrared (MIR) datasets. In typical AGNs, the dusty torus produces strong MIR emission \cite[e.g.,][]{antonucci_1993}, and their MIR colors are redder than those of star-forming (SF) galaxies due to the higher dust temperatures associated with AGN activity \cite[e.g.,][]{donley_2012}. Using {\it Spitzer} and WISE datasets, numerous studies have demonstrated that the MIR color-color diagram is efficient at detecting AGNs, especially those that are dust-obscured \cite[e.g.,][]{lacy_2004, stern_2012,mateos_2012,assef_2018}. Therefore, MIR color-based selection can complement conventional optical methods. 

Additionally, other multiwavelength datasets, such as hard X-ray and radio, which are relatively unaffected by dust obscuration, continue to be essential for obtaining an unbiased census of the AGN population. However, all AGN identification methods suffer from their own selection biases and completeness limits. For instance, hard X-ray AGN selection is ideal for identifying obscured and unobscured AGNs with minimal systematic bias, as high-energy X-ray photons can easily penetrate dense media \cite[e.g.,][]{oh_2018, kim_2021, koss_2022}. Conversely, due to sensitivity limitations, this method is often restricted to relatively nearby sources \cite[e.g.,][]{ho_2001,she_2017}. On the other hand, radio-based selection is biased toward radio-loud sources \cite[e.g.,][]{kellermann_1989,sabater_2019}. Since these comprise only a minor fraction of the total AGN population, this can introduce a bias into the derived statistical properties of the population \cite[e.g.,][]{ho_2001, best_2012}.

As deep MIR photometric data covering wide areas of the sky become increasingly available, the importance of MIR color-based AGN selection has been widely recognized \cite[e.g.,][]{lonsdale_2003,wright_2010, mauduit_2012}. Due to the minimal dust extinction in this wavelength regime, it is considered one of the most effective options for obtaining an unbiased AGN census. However, contamination from starlight or warm dust within SF regions can introduce systematic effects, which often reduce the purity and completeness of the resulting samples \cite[e.g.,][]{lupi_2020, byun_2023, lyu_2024}. Therefore, integrating multiple selection methods is necessary to obtain a complete and unbiased view of AGN demography (\citealp{hickox_2018,siudek_2025}).

Alternatively, the intrinsic variability of AGNs has become a popular method for identification, as time-series photometric data have accumulated significantly through various time-domain surveys such as Stripe 82, ASAS-SN, and ZTF (\citealp{ivezic_2007, kochanek_2017, bellm_2019}). The majority of variability-based AGN selection relies on optical datasets \cite[e.g.,][]{choi_2014,yuk_2022,arevalo_2026}, as the variability characteristics are relatively well known \cite[e.g.,][]{kelly_2009, son_2025}. However, for the MIR variability, its properties are less studied compared to the optical variability \cite[][]{kozlowski_2010, son_2023, kim_2024}. The MIR-based AGN selection has only been applied to a limited number of samples with specific aims \cite[e.g.,][]{pai_2024, 2025ApJ...991...52A}. Consequently, the completeness and purity of MIR-based AGN selection have not yet been systematically validated with a sufficiently large sample. Motivated by this limitation and ongoing MIR missions, such as SPHEREx (\citealp{bock_2025}), this study aims to identify the optimal method for AGN selection based on MIR variability and to evaluate its efficiency and potential systematic biases. In this study, we utilize time-series datasets from the Wide-field Infrared Survey Explorer mission (WISE; \citealp{wright_2010}) for a diverse sample of nearby galaxies. Using these data, we validate the MIR variability-based AGN selection method. Throughout the paper, we adopt a cosmology with $H_0=70$ km s$^{-1}$ Mpc$^{-1}$, $\Omega_m=0.3$, and $\Omega_\lambda=0.7$. 

\begin{figure*}[htp]
\centering
\includegraphics[width=0.97\textwidth]{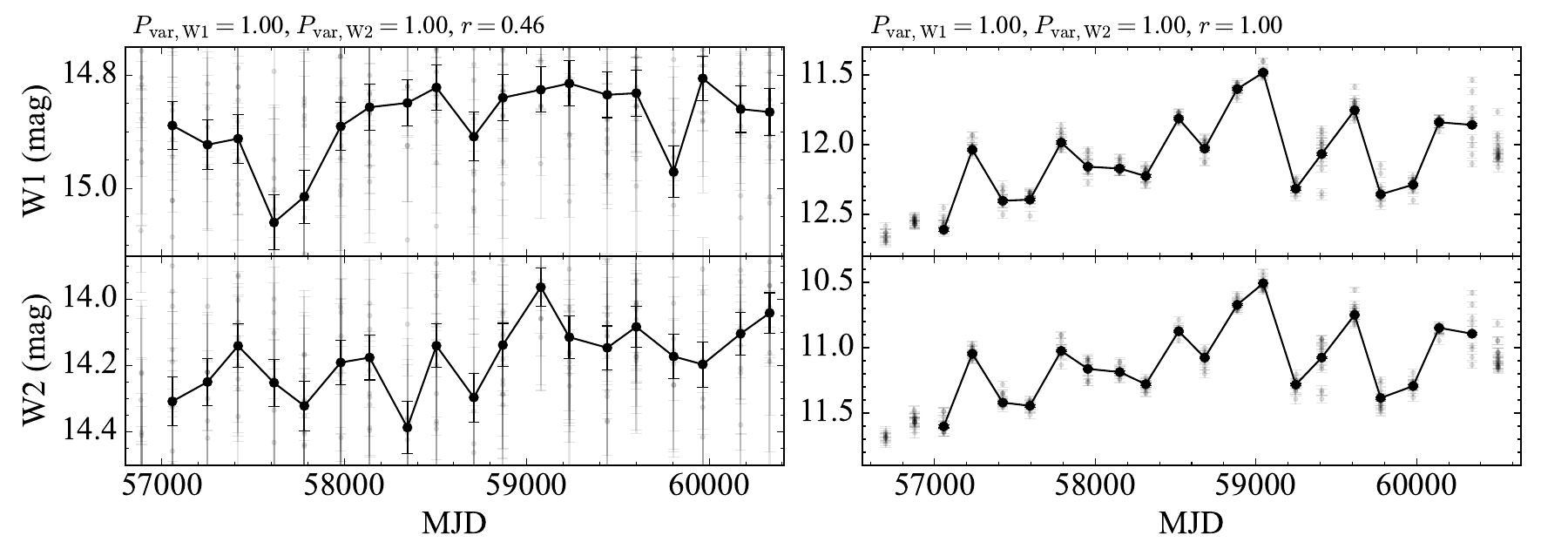}
\caption{Two examples of light curves from our AGN sample. The faint dots denote the original data prior to binning, while the large circles represent the binned data. Due to the artificial offset between AllWISE and NEOWISE, we show only the photometric data from NEOWISE. The top and bottom panels correspond to the W1 and W2 magnitudes, respectively. The left panel shows an example that fulfills the Group 1 criteria but has $r < 0.75$, whereas the right panel shows an example that satisfies the Group 2 criteria.   
}
\end{figure*}

\section{DATA} \label{sec:data}
\subsection{Sample}
To examine the effectiveness of MIR variability in identifying AGNs and to investigate the physical properties of these variability-selected objects, we require a large dataset comprising various galaxy types, including both AGNs and inactive galaxies. To this end, we utilized the Max Planck Institute for Astrophysics and Johns Hopkins University (MPA–JHU) catalog \citep{2004MNRAS.351.1151B}, derived from the SDSS Data Release 8 (SDSS DR8; \citealp{aihara_2011}). 
In this catalog, galaxies with strong emission lines (i.e., S/N $> 3$ for \hb, \oiii, \nii, and \hal) are spectroscopically classified into three subgroups [AGNs\footnote{The majority of the AGN sample are type 2 objects, lacking broad emission lines.}, SF galaxies, and composites] based on the Baldwin, Phillips, \& Terlevich (BPT) diagram (\citealp{baldwin_1981}) combined with the demarcation line adopted from \citet{kewley_2001,kauffmann_2003}. For galaxies with weak emission lines, low S/N subtypes are defined based on the line ratios. For instance, the low-ionization nuclear emission-line regions (LINERs) are defined as galaxies with \nii$/$\hal $> 0.6$, where other emission lines exhibit low S/N ($< 3$). The targets with an \hal\ S/N larger than 2, which have not been classified as AGN or composites, are considered low S/N SF. Finally, galaxies with no detectable emission lines or that cannot be classified using the BPT diagram are classified as normal galaxies. The MPA–JHU catalog includes a total of $\sim1.5$ million galaxies, consisting of 1.8\%, 14.6\%, 59.1\%, 3.5\%, 5.3\%, and 15.7\% classified as AGN, SF, normal, composite, LINER, and low–S/N SF, respectively. It is worthwhile to note that the classification of `LINER' adopted in the MPA–JHU catalog differs slightly from the conventional definition. The classical LINERs are defined as having \oiii/\hb\ $<3$ and \nii/\hal\ $\geq0.6$  \cite[e.g.,][]{ho_1997b}.

We note that duplicate measurements exist in the MPA-JHU catalog. We treat sources within 2 arcsec as duplicates and discard them from the sample. Occasionally, a single target is assigned to multiple subsamples and is therefore excluded from further analysis. This removes $\sim 11.6\%$ of the initial galaxy sample, resulting in a total of $\sim1.3$ million galaxies. The majority of emission-line galaxies exhibit redshifts below 0.4, as the \hb\ and \oiii\ emission lines used for the classification are covered by the SDSS spectral range only up to this redshift. In contrast, normal galaxies span a broader redshift distribution, extending to $z\approx0.7$, which may introduce a potential bias when comparing their MIR variability to that of other galaxy types. 

\subsection{Mid-infrared data} 
We compile the multi-epoch MIR data of the sample from the MPA-JHU catalog, using the AllWISE and NEOWISE data (\citealp{wright_2010, mainzer_2011, AllWISE, neowise}). The AllWISE and NEOWISE scan the entire sky every six months, providing a baseline of $\sim 14$ years. This makes the data well-suited for this study, as the MIR variability timescale in AGNs spans from months to years in the rest-frame \cite[e.g.,][]{son_2023,kim_2024}. We initially cross-match our sample with the WISE dataset with a matching radius of 2 arcsec, considering the point spread function size ($6.1-6.4$ arcsec) of the WISE images. To ensure reliable photometry, we impose the following constraints: \texttt{qual\_frame} $>$ 0, \texttt{qi\_fact }$>$ 0, \texttt{saa\_sep} $>$ 0, \texttt{moon\_masked} = `00', and \texttt{cc\_flags} = `0000' \citep{son_2022}. We utilize $3.4 \mu$m and $4.6 \mu$m band data (W1 and W2, respectively) from AllWISE/NEOWISE, and $12 \mu$m band data (W3) from AllWISE. 

During the survey of the WISE mission survey, typically 14 individual exposures were taken in each visit. Therefore, we bin the multi-epoch data by averaging the photometric results from multiple images using a bin size of 90 days. Upon inspecting the light curves, we occasionally detect outliers in the photometric data within a single-day baseline. These cannot be attributed to intrinsic AGN variability and are likely due to poor PSF fitting, which introduces a substantial bias.

\begin{deluxetable}{lrrrr}[ht]
\tablewidth{0pt}
\footnotesize
\tablecaption{Percentage of MIR variability-based AGNs \label{tab:tab1}}
\tablehead{
\colhead{Optical Class} & \colhead{$N$} & \colhead{Group 1} & \colhead{Group 2} & \colhead{Group 3} \\
\colhead{} & \colhead{} & \colhead{(\%)} & \colhead{(\%)} & \colhead{(\%)} \\
\colhead{(1)} & \colhead{(2)} & \colhead{(3)} & \colhead{(4)} & \colhead{(5)}
}
\startdata
AGNs & 19281 & 34.2 & 28.2 & 9.3 \\
Star-forming & 137092 & 4.6 & 3.3 & 2.1 \\
Normal & 452503 & 2.4 & 1.4 & 1.2 \\
Composite & 39835 & 14.9 & 11.8 & 4.5 \\
LINERs & 60803 & 2.7 & 1.6 & 0.1 \\
Low S/N SF & 163346 & 1.2 & 0.4 & 0.2 \\
\enddata
\tablecomments{
Col. (1): Subsample based on the optical spectral classification.
Col. (2): Subsample size.
Col. (3): Percentage of AGNs selected using the criteria of $P_{\rm var, W1} > 0.99$ and $P_{\rm var, W2} > 0.99$.
Col. (4): Percentage of AGNs selected using the criteria of $P_{\rm var, W1} > 0.99$, $P_{\rm var, W2} > 0.99$, and $r > 0.75$.
Col. (5): Percentage of AGNs selected using the criteria of $P_{\rm var, W1} > 0.99$, $P_{\rm var, W2} > 0.99$, $r > 0.75$, and $P_{\rm AGN,0.8} > 0.5$.
}
\end{deluxetable}

To remove such artificial outliers, we discard photometric data with high reduced $\chi$-square values (i.e., \texttt{w1rchi2} $> 10$ and  \texttt{w2rchi2} $> 10$), as derived from the PSF fit provided by NEOWISE, prior to binning. We also note that the photometric outliers are often associated with large spatial offsets ($> 1$ arcsec). Therefore, we further restrict the matching radius to 1 arcsec. Even with these procedures, some photometric outliers still persist. To minimize systematic uncertainties, we utilize only bins with at least five individual measurements, discard the minimum and maximum values, and perform $3\sigma$ clipping. We carefully examine the resulting light curves and confirm that the majority of outliers are successfully removed with such methods. 

We apply additional corrections to the photometric data to mitigate systematic uncertainties and accurately quantify the light curve variability. We addressed the gradual change in the WISE satellite detector temperature, which results from the decrease in orbital altitude and introduces systematics, by applying the zero-point correction from \citet{son_2026}. Furthermore, a photometric offset was detected between the AllWISE and NEOWISE measurements in the non-variable sources of our sample, which are selected from the normal galaxies using a criterion of $P_{\rm var} < 0.95$ (\S{3.1}). Since this offset could not be accurately corrected using a simple constant, we chose to omit all photometric data from the AllWISE source catalog (\citealp{mainzer_2014}).

Finally, having corrected the NEOWISE dataset with additional treatments, we recalculate the photometric uncertainties based on our processed dataset. For this purpose, we utilize normal galaxies, which are assumed to be non-variable. To exclude variable sources (potentially faint AGNs or transients), we discard sources judged likely to be variable based on their variability probability ($P_{\rm var} \ge 0.95$). To this end, we compute the uncertainties by taking the average of the standard deviations from the light curves of non-variable galaxies within each magnitude bin. We then estimate the photometric uncertainties of individual measurements through interpolation.

Finally, we calculate the variability using the multi-epoch NEOWISE W1 and W2 data, requiring a minimum of 10 epochs per band. This selection results in a final usable sample size of approximately $0.9$ million sources, which corresponds to $\sim59.2\%$ of the initial sample. When considering all classes other than normal galaxies, the average fraction of excluded samples at each step was consistent, ranging from $20\%$ to $40\%$. For the normal galaxies, however, a large number were removed throughout the cleaning process, resulting in an approximate total exclusion of $48\%$. This is likely due to their faintness in the MIR and their higher average redshift compared to the other source types.

\section{Method} \label{sec:method}

To quantify the degree of variability from the NEOWISE dataset and identify AGNs based on the MIR dataset, we adopt three independent parameters: the likelihood of deviation from non-variability ($P_{\rm var}$); the correlation coefficient between W1 and W2 multi-epoch photometry; and the fraction of multi-epoch data that exhibit the AGN-like W1-W2 color ($P_{\rm AGN}$), expressed as a percentage.

\subsection{$P_{\rm var}$}
$P_{\rm var}$ is defined as $1-S(\chi^2, N-1)$, where $S$ is the survival function, $\chi^2$ is the reduced chi-squared statistic calculated from the observed light curve, $N$ is the number of epochs in the light curves of the individual target. $P_{\rm var}$ represents the probability that the observed variance in the light curve is caused by intrinsic variability \citep{mclaughlin_1996}. $P_{\rm var}$ tends to be zero if the fluctuation in the light curves is solely caused by the random measurement errors, whereas $P_{\rm var}$ is close to 1 if its variation is intrinsic and significantly larger than the errors. We calculate $P_{\rm var}$ separately at W1 and W2 bands.    

\begin{figure*}
\centering
\includegraphics[width=0.97\textwidth]{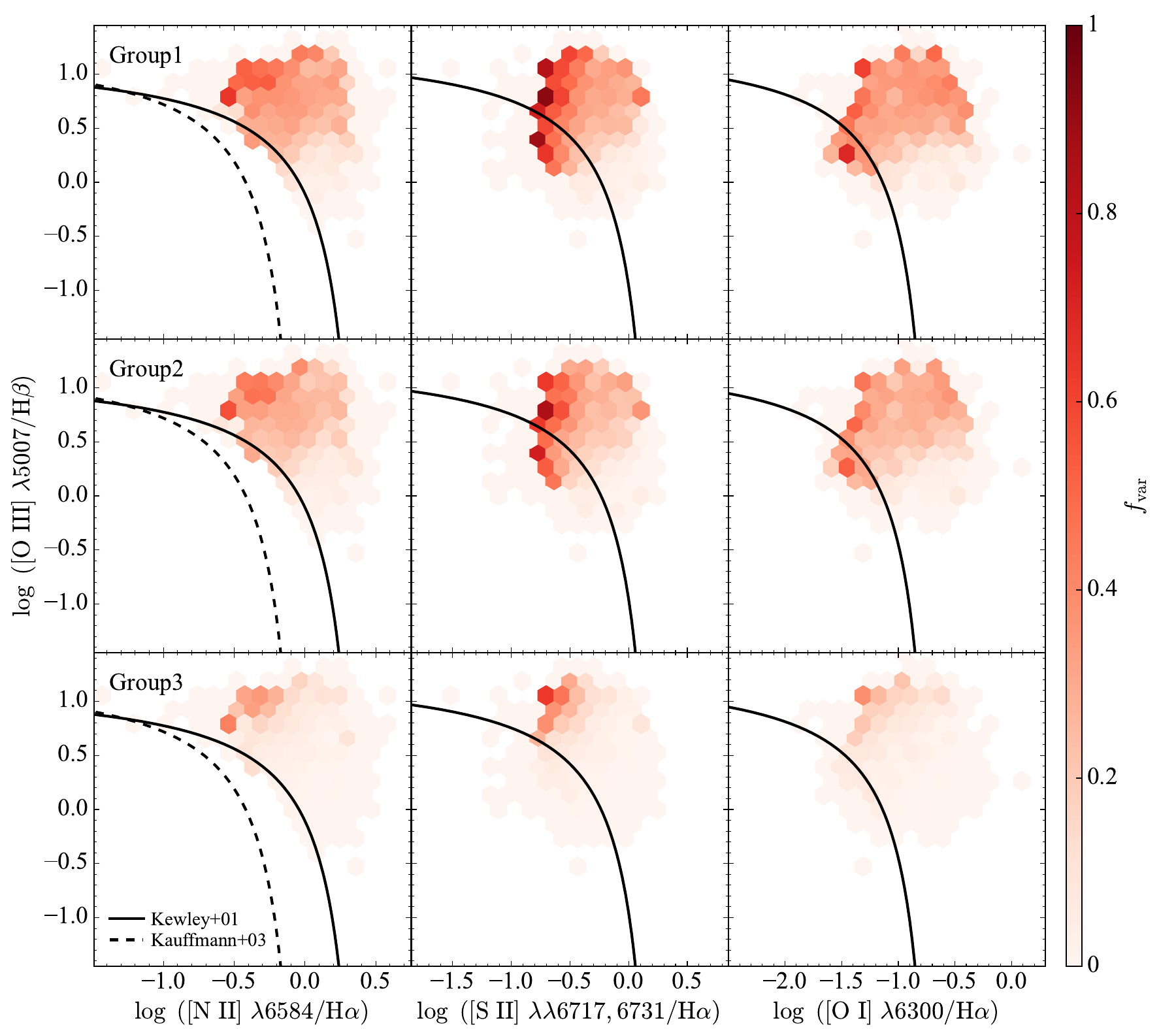}
\caption{The BPT and VO diagrams for the optically selected AGN sample are shown. The fraction of MIR variability–selected AGNs ($f_{\rm var}$) in each bin is indicated by the color maps. The original AGNs were classified as sources above the solid line adopted from \cite{kewley_2001}. SF galaxies were selected below the dashed line \citep{kauffmann_2003}, while sources lying between the dashed and solid lines were classified as composites. From top to bottom, MIR-based AGN classifications are presented in different categories. $f_{\rm var}$ is shown only in bins with more than 10 sources.
}
\end{figure*}

While previous studies adopted a criterion of $P_{\rm var} > 0.95$ to identify variable sources from the MIR light curves \cite[e.g.,][]{sanchez_2017, son_2022}, additional treatments applied to our photometric dataset lead us to set the selection limit in a different manner. After visually inspecting the light curves and their $P_{\rm var}$ values, we find that a threshold of 0.95 includes suspicious sources\footnote{Examples of light curves are shown in the Appendix.}, indicating that a more stringent constraint is required. Therefore, we adopt $P_{\rm var} > 0.99$ for the identification of variable sources throughout this study.

\subsection{Correlation Coefficients}
If the variance of the light curves is primarily caused by random errors, the photometric data for W1 and W2 are unlikely to correlate with each other. Therefore, the correlation coefficient between the W1 and W2 data can be an efficient metric for distinguishing the origin of the variability \cite[e.g.,][]{2025ApJ...991...52A}. To this end, we compute various types of cross-correlation coefficients. Prior to calculating the correlation coefficient, we match the W1 and W2 datasets so that the two light curves share the same time epochs. We use the band with the smaller number of epochs as the reference light curve for this purpose. To synchronize the two datasets from W1 and W2, we only utilize the binned photometric data observed within $\pm$5 days of each other. Throughout this process, a minimum of 8 epochs is granted for our sample.

\begin{figure*}
\centering
\includegraphics[width=0.97\textwidth]{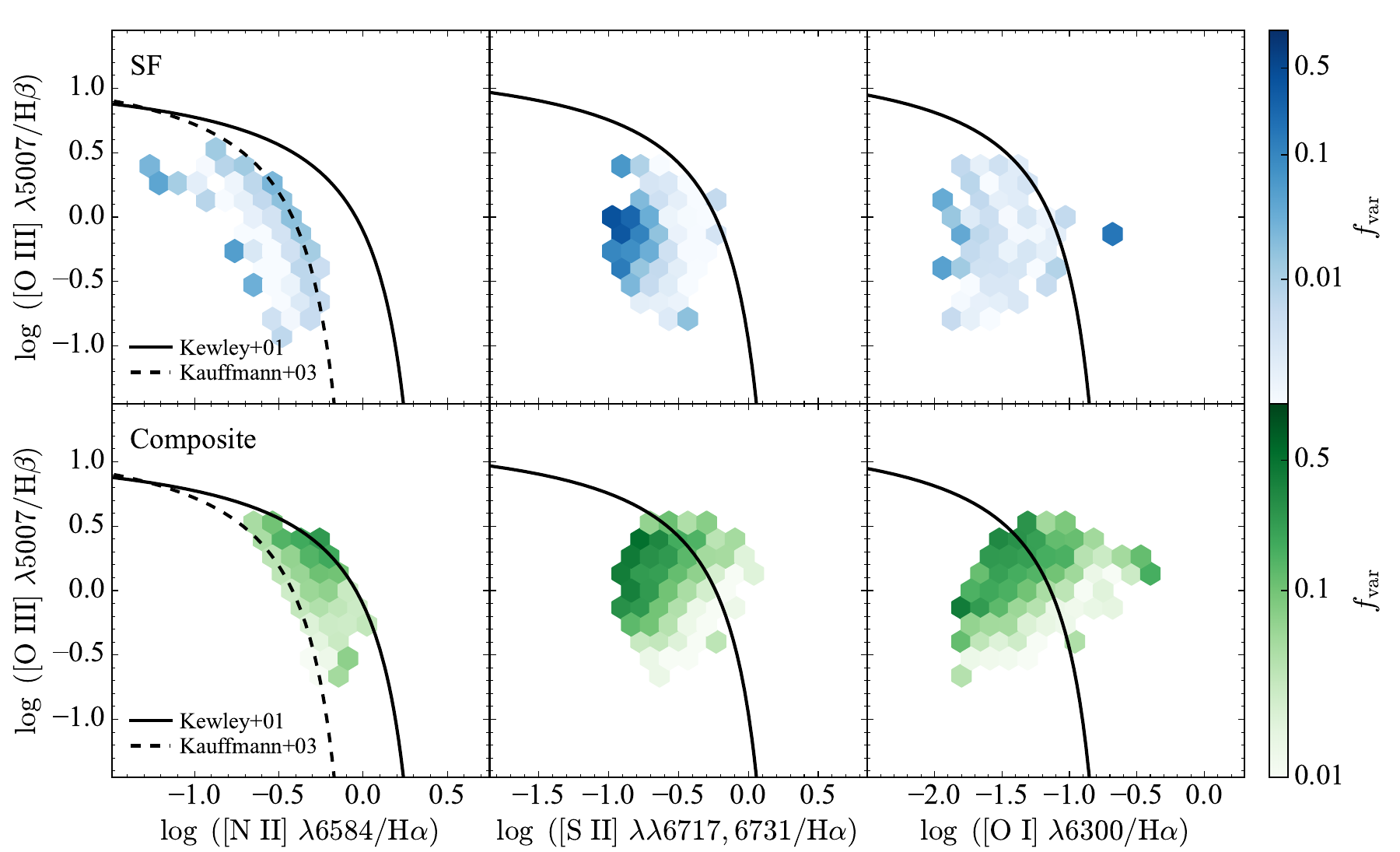}
\caption{ 
Same as Figure 2, except that $f_{\rm var}$ is shown for SF galaxies and composites. The subgroups were originally categorized in the BPT diagram (i.e., \oiii/\hb\ vs. \nii/\hal)}
\end{figure*}

To find the best value for identifying variability, we compute four types of correlation coefficients: the Pearson correlation coefficient ($r$),  the weighted Pearson correlation coefficient ($r_w$), the Spearman rank correlation coefficient ($\rho_S$) and Kendal's tau ($\tau$). For each correlation coefficient, we also calculate the corresponding $p$-value to evaluate its statistical significance. 

We find that all correlation measures exhibit similar behavior, following an approximately Gaussian distribution with an extended tail at the high end. The $r$ distribution specifically exhibits a non-Gaussian profile at $r > 0.75$, as suggested in a previous study \citep{2025ApJ...991...52A}. Consequently, we select $r$ to identify variable sources using a threshold of 0.75. To avoid selecting statistically insignificant sources, we impose an additional criterion of $p < 0.05$.

\subsection{$P_{\rm AGN}$}
It is well established that the higher dust temperatures in the AGN torus, compared to the relatively cold dust typically found in SF galaxies, produce distinctive MIR colors that separate AGNs from inactive galaxies \cite[e.g.,][]{lacy_2004,stern_2005,assef_2018}. Motivated by this, \citet{2025ApJ...991...52A} showed that the W1–W2 color is an effective diagnostic for identifying AGNs, complementing MIR variability, as expected from the AGN wedge in MIR color space. Following their approach, we adopt $P_{\rm AGN}$, defined as the fraction of the observing period during which the matched W1–W2 color exceeds a specified threshold. We compute $P_{\rm AGN}$ for the threshold: W1-W2 $\ge 0.8$, which we denote as $P_{\rm AGN,0.8}$ \citep{stern_2012}. However, since $P_{\rm AGN}$ is not directly linked to intrinsic AGN variability, we employ it as a supplementary diagnostic for validating our variability-based AGN selection method throughout this study.
\begin{figure*}[tp]
\centering
\includegraphics[width=0.97\textwidth]{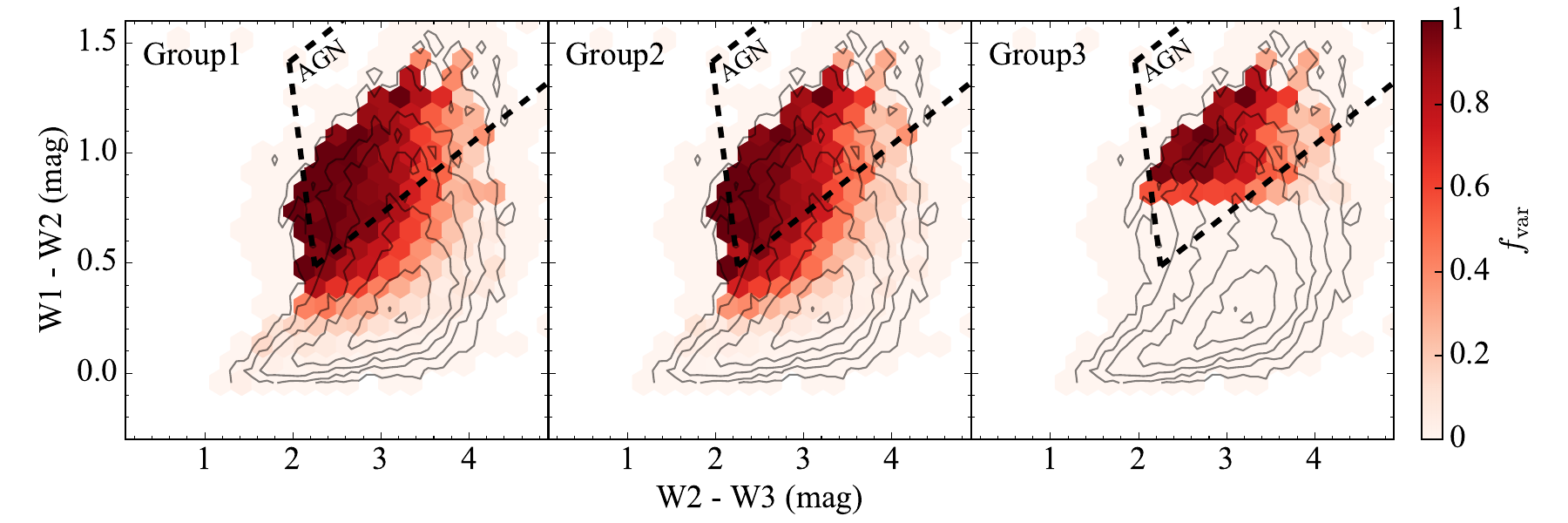}
\caption{ 
Distributions of MIR variability–based AGN fractions in the W1–W2 versus W2–W3 color diagram. The contours represent the density distribution of the parent sample, while the color map shows the fraction of AGN sources in each bin, categorized according to the three different criteria. The AGN fraction is only demonstrated in bins containing more than 10 sources. The thick dashed lines show the AGN wedge adopted from \cite{mateos_2012}.
}
\end{figure*}

\section{RESULTS} \label{sec:results}

\subsection{Fiducial Test with AGNs}
To validate the MIR variability-based AGN selection method, we first apply this method to optically selected AGNs. 
Using the three previously described methods, we classified galaxies into variability Groups: 

\begin{description}
    \item[Group 1] $P_{\rm {var}} > 0.99$
    \item[Group 2] $P_{\rm {var}} > 0.99 \ \& \ r>0.75 $
    \item[Group 3] $P_{\rm {var}}> 0.99 \ \& \ r>0.75 \ \& \ P_{\rm {AGN,0.8 }} > 0.5$
\end{description}

Of the entire sample of AGNs, $\sim34\%$ exhibit variability based on the criteria of Group 1, in which $P_{\rm var}$ exceeds 0.99 for both bands. In general, $P_{\rm var}$ alone is sufficient to identify variable sources. However, instrumental failures, such as temporal fluctuations, can introduce artificial variation in the light curves. This process can result in a significant overestimation of $P_{\rm var}$, leading to spurious variability detections. Additionally, the same impact can occur when the measurement error is underestimated. Consequently, the resulting purity of Group 1 can be low. This effect is further discussed in \S{5.1}.

The variability fraction defined by Group 2 in the AGNs is marginally decreased to $\sim28\%$, with an additional criterion of $r > 0.75$ from Group 1. While $P_{\rm var}$ is estimated in the individual band, the correlation coefficient captures simultaneous changes in both bands over the observation baseline, which is suitable for effectively detecting variance due to non-random noise. Therefore, with this additional constraint, some sources regarded as variable in Group 1, possibly due to the outliers in the photometric data, can be properly removed. 

Examples of light curves are shown in Figure 1, including one object satisfying Group 1 criteria (but not Group 2) and one object classified as Group 2. An object with high $P_{\rm var}$ but low $r$ indicates that variability is detected independently in each band, while the W1 and W2 bands remain uncorrelated. In such cases, the source's variability is unlikely to originate from intrinsic physical changes. Instead, variations arise independently in each band, possibly due to random fluctuations, leading to a lower $r$ value. We note that relaxing the constraint to $P_{\rm var} > 0.95$ in both bands increases the Group 2 variability fraction by only $0.5\%$, suggesting that the $P_{\rm var}$ criterion is not a dominant factor in our analysis.

Notably, when applying the MIR color cut for Group 3, the MIR-selected AGN fraction dramatically decreases to $\sim9\%$. This suggests that color-based selection may lead to the exclusion of a significant number of AGNs, particularly low-$z$ AGNs. This trend is attributed to the fact that the MIR emission is heavily affected by the stellar light from the host galaxy and dust emission from SF regions \cite[e.g.,][]{ciesla_2015, son_2022}. Previous studies also demonstrated that a significant fraction of type 1 and 2 QSOs do not fulfill the MIR color criteria (i.e., W1$-$W2 $< 0.8$; e.g., \citealt{byun_2023}). \citet{son_2022} showed that the W1$-$W2 color is strongly dependent on the AGN luminosity, as the relative contribution from host galaxies is anti-correlated with the AGN luminosity. Given that our AGN sample is predominantly composed of moderate-luminosity AGNs, the MIR color cut may be inadequate for identifying such AGNs. Consequently, in this study, we identify AGN candidates based on MIR variability using the criteria of Group 2.

\subsection{MIR Variability in Inactive Galaxies}
We identified AGN candidates in other types of galaxies using the selection criteria (Group 2) derived from the fiducial test with optically selected AGNs. Among optically selected SF galaxies, $3.3 \%$ of sources can be categorized as AGNs, suggesting that emission lines originating from AGNs may be weak (possibly due to dust obscuration) or be overwhelmed by the emission from intense SF. Interestingly, when applying the MIR color cut, the fraction of AGNs only marginally decreases to $2.1 \%$, implying that the detected MIR variability mostly arises from AGNs.

\begin{figure}
\includegraphics[width=0.49\textwidth]{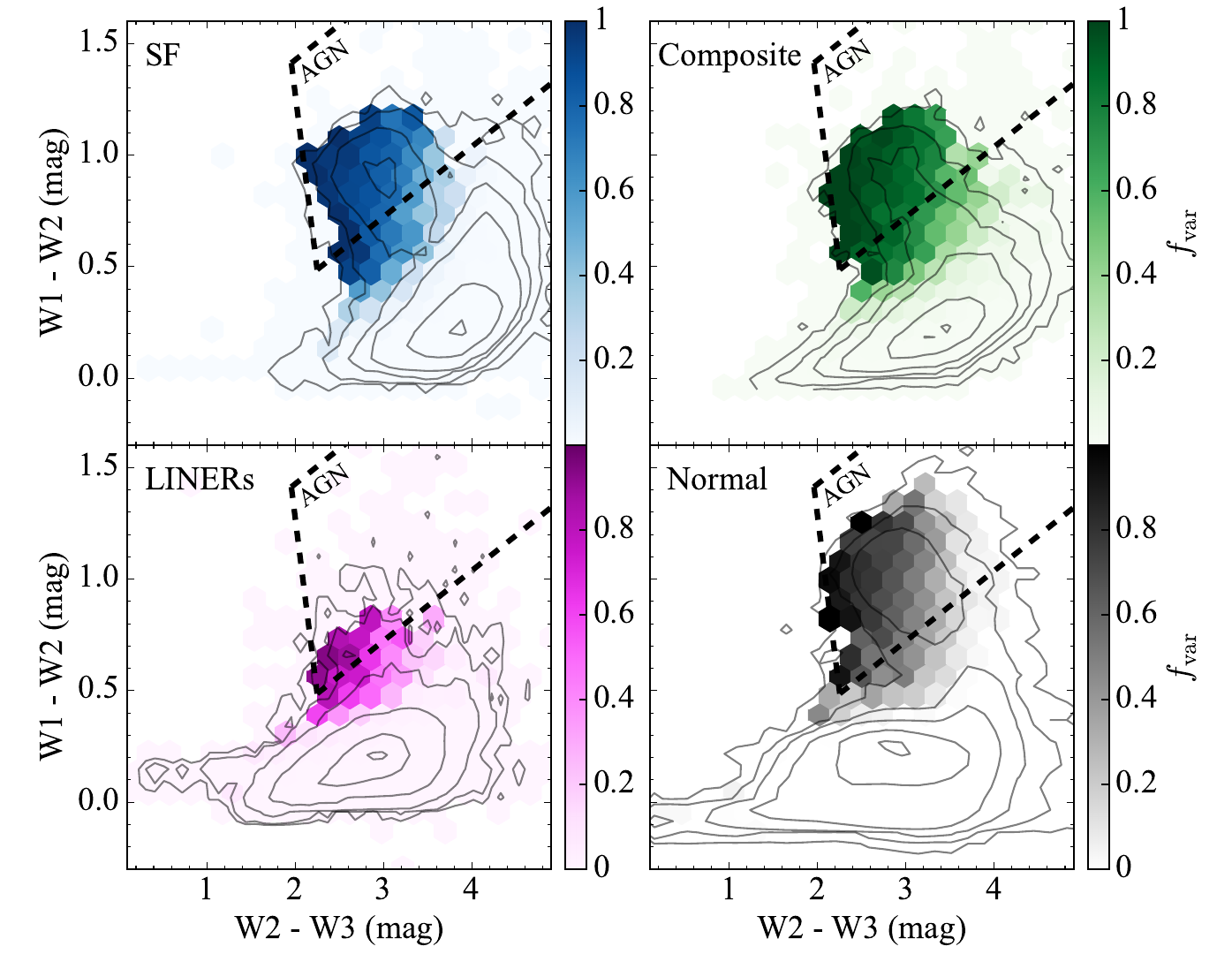}
\caption{
Same as Figure 4, for different subsamples. 
}
\end{figure}

Conversely, the AGN fraction for composites ($11.8\%$) is substantially larger than that for SF galaxies. This is expected, as composites are located between SF galaxies and AGNs in the BPT diagram. The connection between the MIR variability and the location at the BPT diagram is further discussed in Section \ref{sec:discussion}. For galaxies with weak or no emission lines (normal, LINER, and low S/N SF galaxies), the AGN fraction derived from the MIR variability is almost negligible, ranging from 0.4 to $1.6\%$. Although this is not surprising given the weakness of the emission lines, it is somewhat unexpected that these three classes share a similarly low fraction, since LINERs can be powered by weak AGNs \cite[e.g.,][]{ho_1993}.
It is worthwhile noting that, due to the detection limit of line emission and the limited wavelength coverage of SDSS spectra, normal galaxies exhibit higher redshift ranges compared to other subgroups. When restricting the redshift of normal galaxies to 0.4, the AGN fraction drops significantly to 0.14\%. The detailed AGN fractions of subgroups based on the different selection criteria are summarized in Table 1.   

\section{DISCUSSION} \label{sec:discussion}

\subsection{Comparison with Other AGN Diagnostics}
Conventionally, the classification of obscured AGNs has been carried out using various types of diagnostics, including line ratios of narrow emissions, MIR colors, and radio loudness \cite[e.g.,][]{veilleux_1987, stern_2012, ivezic_2002}. Each method has its own advantages and disadvantages. For example, the line-ratio-based method is suitable for application to large spectroscopic datasets; however, highly obscured AGNs, AGNs with prominent SF, or AGNs with weak emission lines can be missed \cite[e.g.,][]{agostino_2019}. On the other hand, AGN selection based on MIR colors is effective for searching for obscured AGNs, but contamination from the host galaxies and SF regions may lead to a reduction in the completeness of the AGN sample \cite[e.g.,][]{lacy_2004}. To properly examine the characteristics of the MIR-variability-based method, comparisons with previous AGN selection methods are essential.

The MPA-JHU catalog originally categorized galaxies based on their location on the BPT diagram. Therefore, it is worthwhile to compare our result with the one in that diagram. Figures 2 and 3 show the distribution of AGN fraction determined by the MIR variability in each bin. The fraction of MIR-selected AGNs is largest in the AGN location, gradually decreases in the composite region, and drops rapidly in the SF region, as expected. This clearly demonstrates the reliability of our method.  

\begin{figure}
\includegraphics[width=0.49\textwidth]{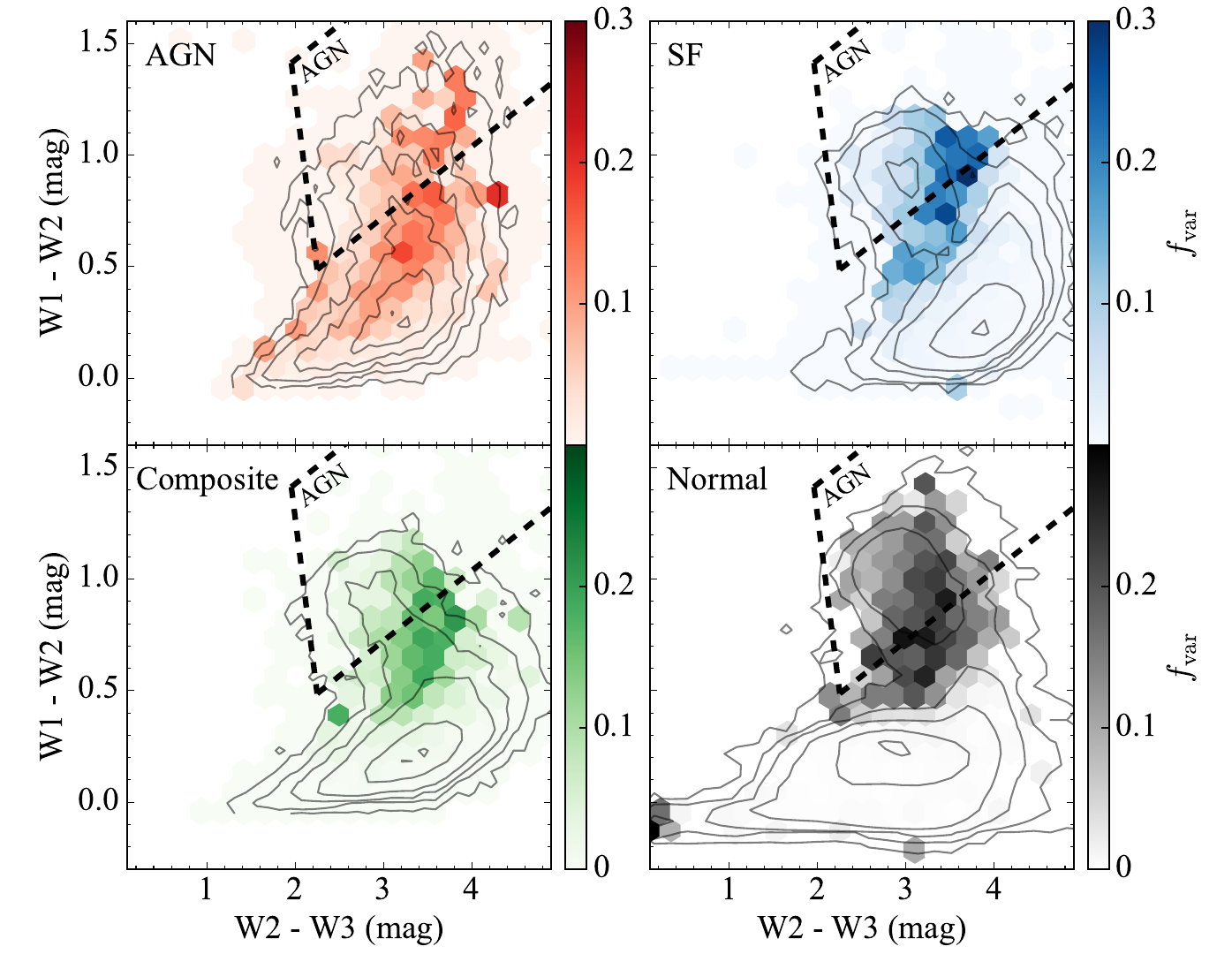}
\caption{
Same as Figure 4, for optically selected AGNs categorized as Group 1 but fulfilling Group 2 criteria. 
}
\end{figure}

MIR color–color diagrams constructed from observations with the {\it Spitzer} telescope and the WISE survey have been widely employed to identify AGNs, particularly those that are dust-obscured \cite[e.g.,][]{lacy_2004, stern_2012}. For our comparison, we adopt representative W1 and W2 magnitudes computed as weighted means of the multi-epoch measurements for each source, while the W3 magnitudes are taken from the AllWISE catalog. We note that a subset of the sources is not detected in the W3 band and is therefore excluded from subsequent analysis. The W3 non-detection fraction varies significantly among the different subgroups, being lowest for AGNs and highest for normal galaxies, primarily due to the intrinsic strength of MIR emission originating from warm dust.

\begin{deluxetable}{lrrrrr}[ht!]
\tablewidth{0pt}
\footnotesize
\tablecaption{Percentage of MIR variability-based AGNs within the MIR AGN Wedge \label{tab:tab2}}
\tablehead{
\colhead{Opitcal Class} & \colhead{\h $N$} & \colhead{\h $f_{\rm wedge}$} & \colhead{\h Group 1} & \colhead{\h Group 2} & \colhead{\h Group 3} \\
\colhead{} & \colhead{} & \colhead{(\%)} & \colhead{(\%)} & \colhead{(\%)} & \colhead{(\%)} \\
\colhead{(1)} & \colhead{(2)} & \colhead{(3)} & \colhead{(4)} & \colhead{(5)} & \colhead{(6)}
}
\startdata
AGNs &\h 18030 &\h 15.5 &\h 37.1 &\h 42.0 &\h 90.6 \\
Star-forming &\h 127055 &\h 3.1 &\h 58.4 &\h 71.7 &\h 97.1 \\
Normal &\h 256401 &\h 3.6 &\h 74.5 &\h 88.9 &\h 98.3 \\
Composite &\h 37115 &\h 6.4 & 38.8 &\h 45.7 &\h 93.6 \\
LINERs &\h 51615 &\h 0.5 &\h 13.4 &\h 19.2 &\h 93.2 \\
Low S/N SF &\h 132856 &\h 0.3 &\h 23.4 &\h 52.8 &\h 93.6 \\
\enddata
\tablecomments{ 
Col. (1): Subsample based on the optical spectral classification.
Col. (2): Number of sources in the subsample with reliable photometric measurements in W1, W2, and W3.
Col. (3): Percentage of sources lying within the MIR AGN wedge adopted from \cite{mateos_2012}.
Col. (4): Percentage of Group 1 sources satisfying the MIR AGN wedge selection criteria.
Col. (5): Percentage of Group 2 sources satisfying the MIR AGN wedge selection criteria.
Col. (6): Percentage of Group 3 sources satisfying the MIR AGN wedge selection criteria.
}
\end{deluxetable}

Figure 4 shows the fraction of variable sources, categorized by different Group criteria, within the W1$-$W2 vs. W2$-$W3 diagram for optically selected AGNs. It is evident that the majority of AGNs selected based on MIR variability satisfy the AGN wedge adopted from \citet{mateos_2012}, and the fraction of AGNs sharply decreases as the MIR colors deviate from this wedge. The selection based on Group 3 criteria significantly rejects genuine AGNs, thereby reducing the completeness of this method. More importantly, a similar trend is observed in other galaxy types, indicating the reliability of our MIR variability-based selection method (Fig. 5). Table 2 lists the fraction of MIR variability-selected AGNs in each subsample located within the AGN wedge. To indirectly assess the purity of Group 1 criteria, we examine the fraction of variable sources that satisfy Group 1 but do not fulfill Group 2 criteria in the MIR color-color diagrams (Fig. 6). Contrary to variable sources chosen by ordinary selection criteria (i.e., Groups 1, 2, and 3), these objects are widely distributed across the color-color diagrams, suggesting that false detections may be moderate.

\begin{figure}
\includegraphics[width=0.48\textwidth]{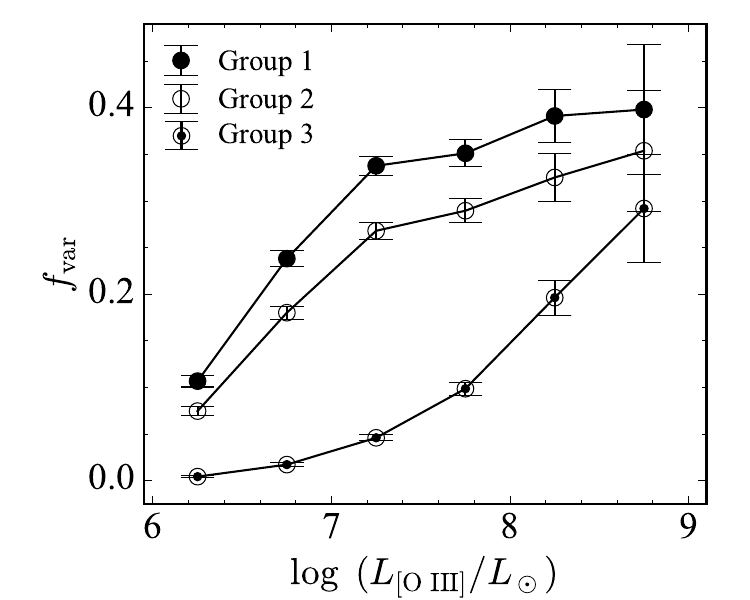}
\caption{The fraction of MIR-based AGNs among optically selected AGNs as a function of \oiii\ luminosity. Subgroups based on different selection criteria are shown separately. The uncertainties are computed assuming Poisson statistics.} 
\end{figure}

Furthermore, although incomplete, the radio emission can serve as an indicator of AGN activity. Therefore, it can be used to examine the reliability of our MIR-based selection. To this end, we cross-match our sample with Epoch 1 \citep{gordon_2020} of the Very Large Array Sky Survey (VLASS) radio data \citep{lacy_2020}, which provides continuum data at $2-4$ GHz with a $1\sigma$ sensitivity of $\sim120\ \mu$Jy over the sky area $\delta > -40^{\circ}$. Using a matching radius of 2\asec, we find that $\approx 8.0\%$, $0.3\%$, $2.7\%$, and $2.1\%$ of the parent sample have counterparts in the VLASS for AGNs, SF, composites, and normal galaxies, respectively. Notably, the fraction of radio detections increases by a factor of $7-8$ ($1.9\%$ for SF galaxies and $16.5\%$ for normal galaxies) for AGN candidates (Group 2). This indirectly demonstrates the reliability of our MIR-based AGN selection. However, such an increase in radio detection among AGN candidates is not observed in AGNs or composites. While the physical origin of the discrepancy between SF/normal galaxies and AGNs/composites is unclear, this may partially stem from the inverse correlation between radio loudness and Eddington ratio \citep{ho_2002}.

\begin{figure*}
\hspace{-0.4cm}
\includegraphics[width=0.37\textwidth]{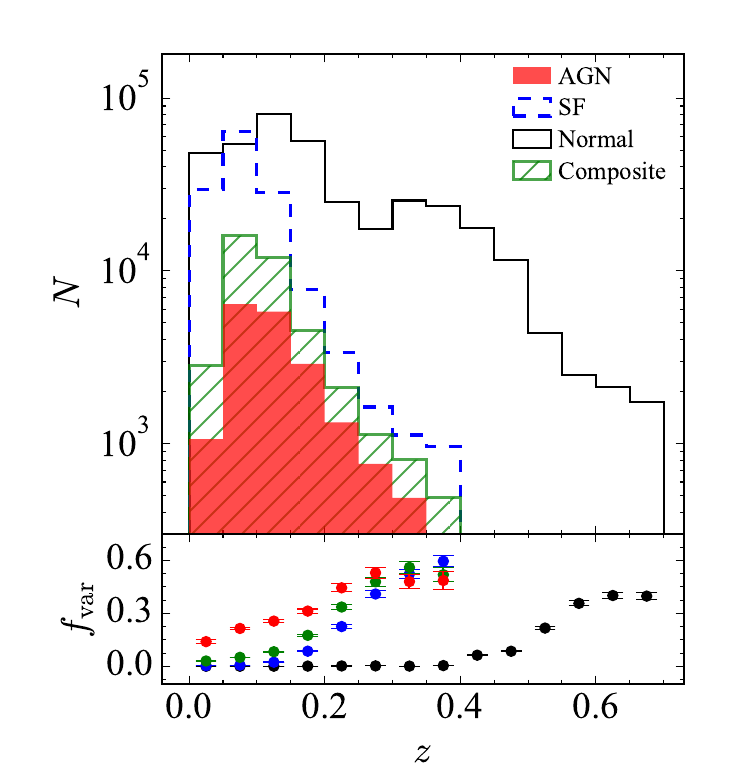}
\hspace{-0.7cm}
\includegraphics[width=0.37\textwidth]{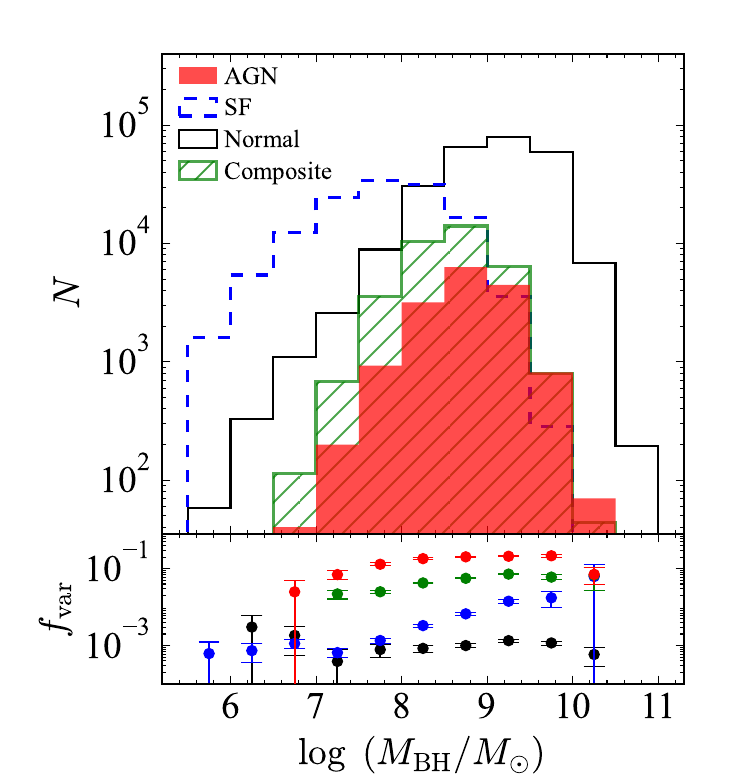}
\hspace{-0.7cm}
\includegraphics[width=0.37\textwidth]{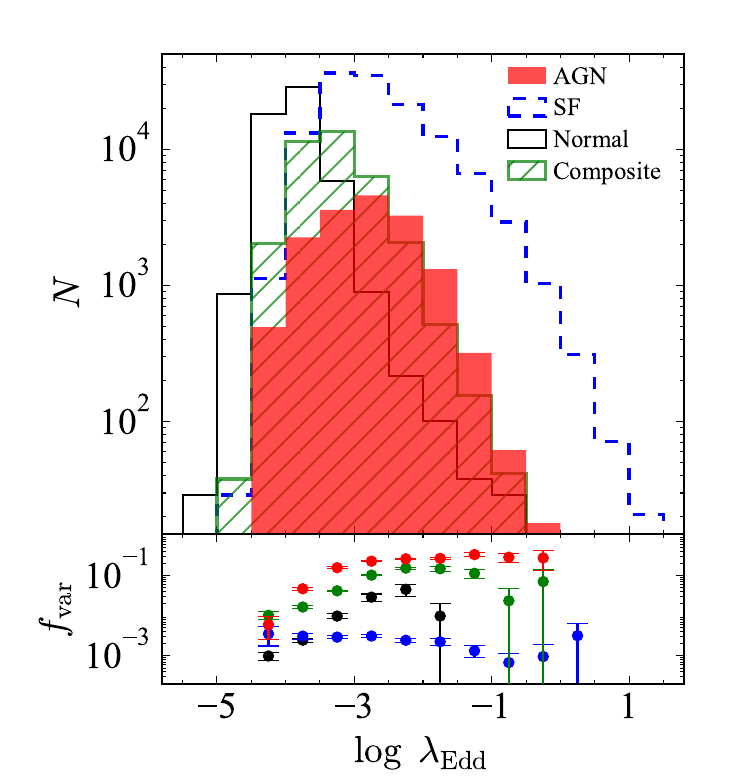}
\caption{ Distribution of the sample properties and MIR variability fractions. The top panels show the source counts for the four spectral subgroups, while the bottom panels show the corresponding fraction of MIR variable sources. From left to right, the distributions are plotted as a function of redshift ($z$), black hole mass ( $M_{\text{BH}}$), and Eddington ratio ($\lambda_{\text{Edd}}$).} 
\end{figure*}

\subsection{Characteristics of MIR Variability-selected AGNs}
Because all AGN selection techniques are subject to their own systematic effects, assessing the implications of our selection methodology is warranted. To this end, we examine how the AGN fraction depends on the physical properties of the AGNs. We first find that the fraction of MIR variability-based AGNs ($f_{\rm var}$) increases with the AGN brightness, as traced by the \oiii\ luminosity in optically selected AGNs (Fig. 7).  
Furthermore, Figure 8 illustrates $f_{\rm var}$ as a function of redshift, BH mass ($M_{\rm BH}$), and Eddington ratio ($\lambda_{\rm Edd}$) across four distinct subgroups categorized by SDSS optical spectra. $M_{\rm BH}$ is estimated from the stellar mass using the $M_{\ast}$–$M_{\rm BH}$ relation (for all galaxy types) from \citet{greene_2020}, while the bolometric luminosity ($L_{\rm bol}$) is derived from the \oiii\ luminosity using the bolometric correction from \citet{heckman_2005}. We assume that the entire \oiii\ flux originates from the AGN instead of star formation, which is a reasonable approximation for relatively massive, and hence metal-rich, host galaxies \citep{ho_2005}. Nevertheless, the Eddington ratios for non-AGN subgroups should be regarded as upper limits. We note that even though we adopt other bolometric corrections, for example, those from \citep{lamastra_2009}, which account for extinction based on the Balmer decrement, the main result remains unchanged (see \citealp{kong_2018} for extensive discussion).

Notably, the fraction of variable sources increases dramatically with redshift across all subgroups. While the primary driver of this trend remains unclear, a combination of two factors (AGN luminosity and the central wavelengths of the WISE filters) may play a key role. Due to Malmquist bias, distant objects in our sample tend to be more luminous than nearby ones. Furthermore, as redshift increases, the central wavelengths of the W1 and W2 bands shift toward the peak of hot dust emission ($\sim 2$--$3$ $\mu$m; \citealp{mor_2011}), making the observed fluxes more sensitive to MIR variability. We also find that variable sources are more common at higher BH masses. Because AGN luminosity is proportional to BH mass for a constant Eddington ratio, the observed BH mass dependence likely reflects the underlying relationship with AGN luminosity. Finally, the fraction of variable sources appears to peak at moderate Eddington ratios and decline at the extreme low end of the distribution. Interestingly, this finding is consistent with theoretical and observational studies suggesting that the torus fails to form at low Eddington ratios \cite[e.g.,][see \S{5.3} for further discussion]{elitzur_2006}. On the other hand, the torus may disappear at high Eddington ratios \cite[e.g.,][]{ho_2012, venanzi_2020}, where a decrease in $f_{\rm var}$ is predicted. The limited sample size of our dataset in this regime prevents us from testing this scenario.

As our method, based on MIR variability, is effective in identifying AGN candidates in dust-obscured sources and SF-dominated galaxies, it may be ideal to examine the connection between AGNs and the physical properties of galaxies, such as the star formation rate (SFR). With this motivation, we compare the locations of AGNs identified in this study in the diagram of stellar mass and SFR, which provides useful insights into the coevolution between SMBHs and their host galaxies. Note that stellar masses were calculated by combining the mass-to-light ratio at the $z$-band, derived from the modeling of observed spectra, and the $z$-band luminosity (\citealp{2003MNRAS.341...33K}). 
The SFRs were estimated using the emission lines for those with detected \hal\ emissions (\citealp{2004MNRAS.351.1151B}). In contrast, SED fitting to the SDSS photometry was utilized to compute the SFRs of galaxies exhibiting no nebular emission lines (\citealp{salim_2007}).

\begin{figure} 
\centering
\includegraphics[width=0.48\textwidth]{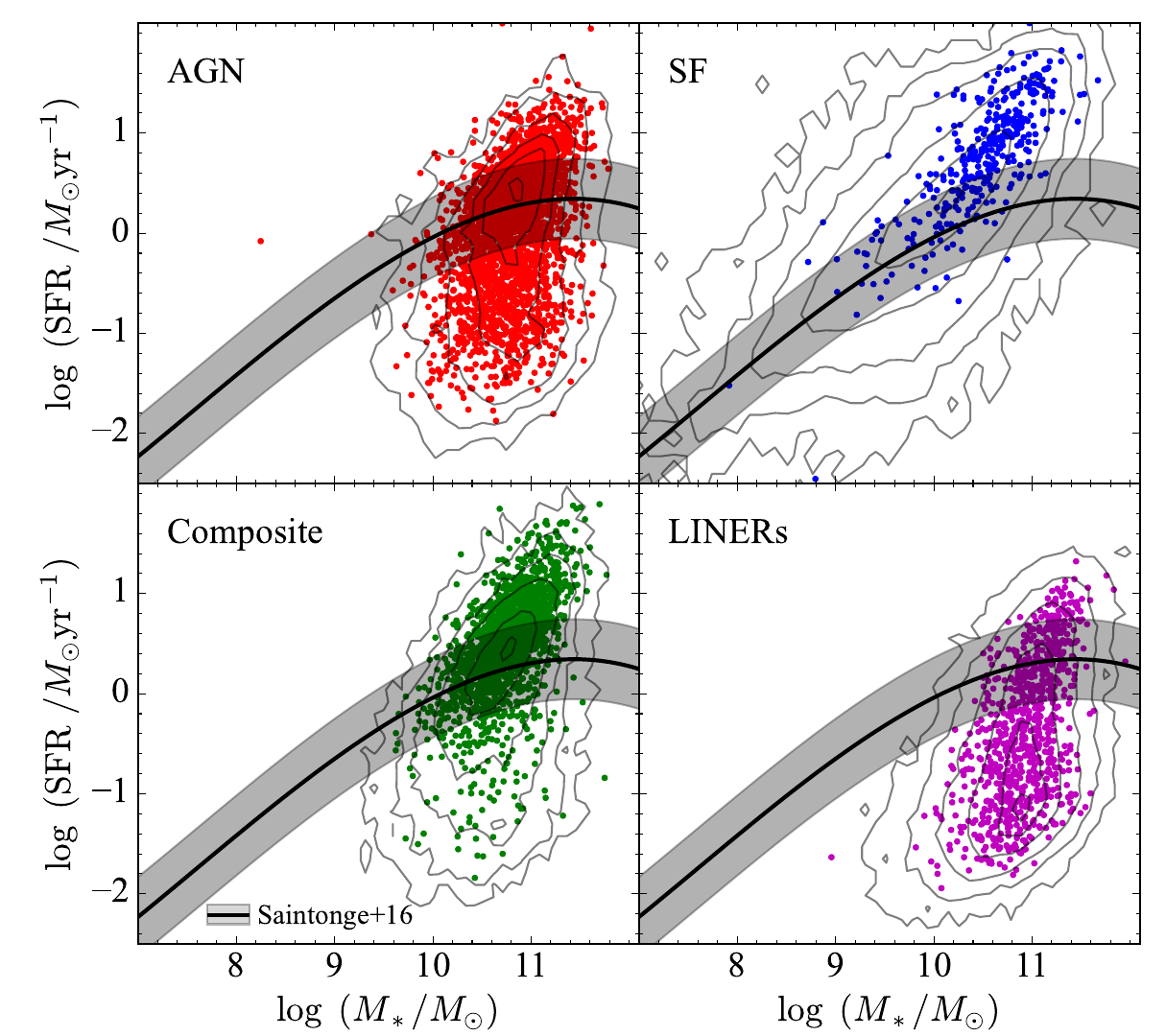}
\caption{The relation between stellar mass and star formation rate for different subgroups. The background contours represent the parent sample in each subgroup on logarithmic scales, while the filled circles denote the MIR variability–selected AGNs based on the Group 2 criteria. The solid line indicates the star formation main sequence adopted from \cite{saintonge_2016}. The gray-shaded region denotes the $\pm0.4$ dex, $1\sigma$ scatter for SFMS.}
\end{figure}

Figure 9 shows the distribution of subsamples (including MIR variability-based selected AGNs) in the stellar mass-SFR diagram. SF galaxies are known to follow the relation between the star SFR and the stellar mass, which is the star formation main sequence (SFMS; \citealp{noeske_2007, renzini_2015, li_2023}). To investigate the physical connection between SF and AGN activity, it is beneficial to account for this relation. Therefore, we adopt the deviation from the SFMS, which is defined as ($\Delta {\rm SFMS} \equiv \log ({\rm SFR/SFR_{\rm MS}})$), where ${\rm SFR_{MS}}$ is the SFR of the SFMS at a given stellar mass adopted from \citet{saintonge_2016}. Figure 10 shows the relation between $\Delta {\rm SFMS}$ and the fraction of AGN at a given stellar mass/SFR bin. Notably, the two quantities exhibit a positive correlation across all subgroups, suggesting that AGN activity is synchronized with the SF. Furthermore, the observed increase of $f_{\rm var}$ with the AGN brightness (Fig. 7) suggests a correlation between AGN luminosity and SFR. Interestingly, this finding is consistent with previous studies based on different techniques, such as the photometric properties of host galaxies, spectral features of young stars, and the FIR-based SFR in AGNs \cite[e.g.,][]{kauffmann_2003, diamond_2012, kim_2019, zhuang_2023}. This clearly demonstrates the power of our AGN selection method based on the MIR variability, as it allows us to detect AGNs in various types of galaxies, including SF-dominated galaxies and composite systems, from a large survey dataset.

\begin{figure} 
\includegraphics[width=0.48\textwidth]{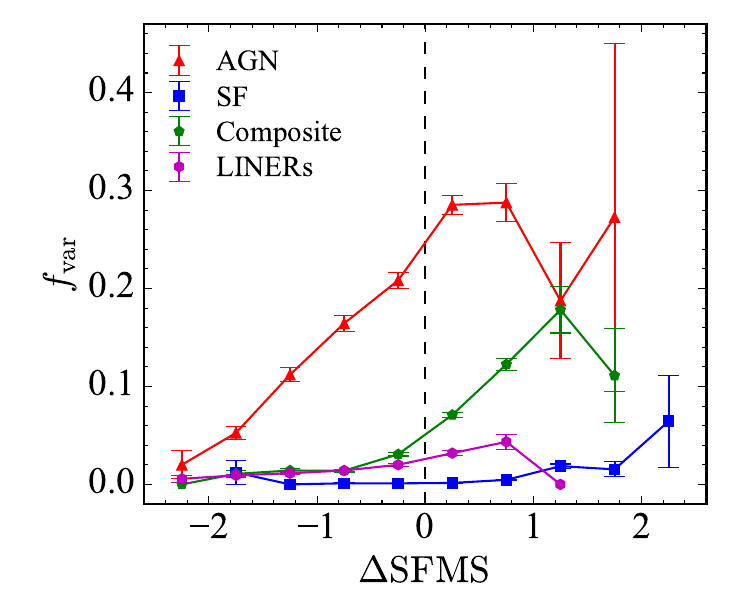}
\caption{The fraction of MIR-selected AGNs as a function of the deviation from the SFMS ($\Delta{\rm SFMS}$) at a given stellar mass. Red, blue, green, and purple symbols represent the fractions for AGNs, SF galaxies, composites, and LINERs, respectively.} 
\end{figure}

\subsection{Implication for Classical LINER}
The nature of LINERs, as studied through integrated, global spectroscopy, can be ambiguous because the large aperture mixes various excitation mechanisms (AGNs, shocks, and old stars).  However, when properly isolated with high-resolution observations, and especially when combined with multiwavelength information, LINERs are AGNs of low Eddington ratio (\citealp{ho_2008, ho_2009}). Here, we utilize the MIR variability of LINERs to investigate their physical properties. For clarification, we adopt the conventional definition of LINERs, which is distinguished from the one employed in the MPA-JHU catalog. Specifically, we classify classical LINERs located below the demarcation line from \citet{ho_1997b} on the BPT diagram, resulting in 7433 objects ($\sim 46.1 \%$ of optically selected AGNs). Note that LINERs from the MPA-JHU catalog are not used in the analysis due to their distinctive definition and low S/N of emission lines.  

The fraction of sources associated with MIR variability significantly decreases in the classical LINERs to $\sim 11.5\%$ (Fig. 2). Furthermore, unlike the composites and SF galaxies, the classical LINERs in the MIR color-color diagram are mostly located outside the AGN wedge (Fig. 11). The fraction of sources within the AGN wedge is even smaller than that of normal galaxies. This result may suggest that the classical LINERs possess low-luminosity AGNs, where their MIR color is mostly dominated by the stellar light, or that they have a non-AGN origin \cite[][]{ho_1993,ho_2008,ho_2009}. However, these explanations are insufficient to account for the extremely low MIR variability and the redder MIR colors observed in these galaxies compared to other inactive galaxies (see also \citealp{ho_2009}).

Instead, they may be low-luminosity AGNs that lack the dusty torus, unlike ordinary luminous AGNs. This is the prediction from the disk-wind driven torus formation, where the outflow from the accretion disk drives the formation of the dusty torus and dense gas clouds in the broad line regions \cite[e.g.,][]{elitzur_2006,elitzur_2009,wada_2012}. In this model, the broad line region and torus may disappear in low-luminosity AGNs because the outflow cannot be sustained under these conditions. Notably, this prediction is adequate to explain the extremely low MIR luminosity of the classical LINERs (see also \citealp{ho_1999,ho_2008,ho_2009}).      

\begin{figure}
\centering
\includegraphics[width=0.48\textwidth]{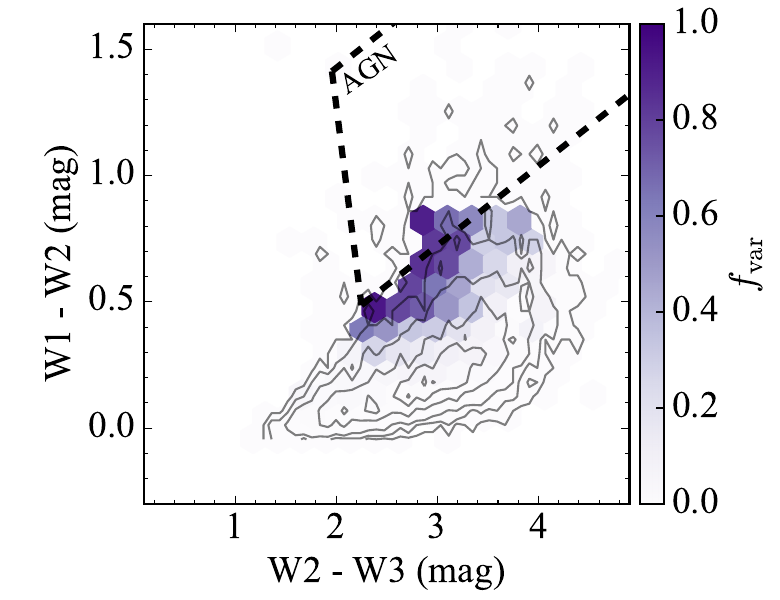}
\caption{Same as Figure 4, for classic LINERs classified using the definition from \cite{ho_1997b}.} 
\end{figure}

\begin{figure*}[htp]
\centering
\includegraphics[width=0.97\textwidth]{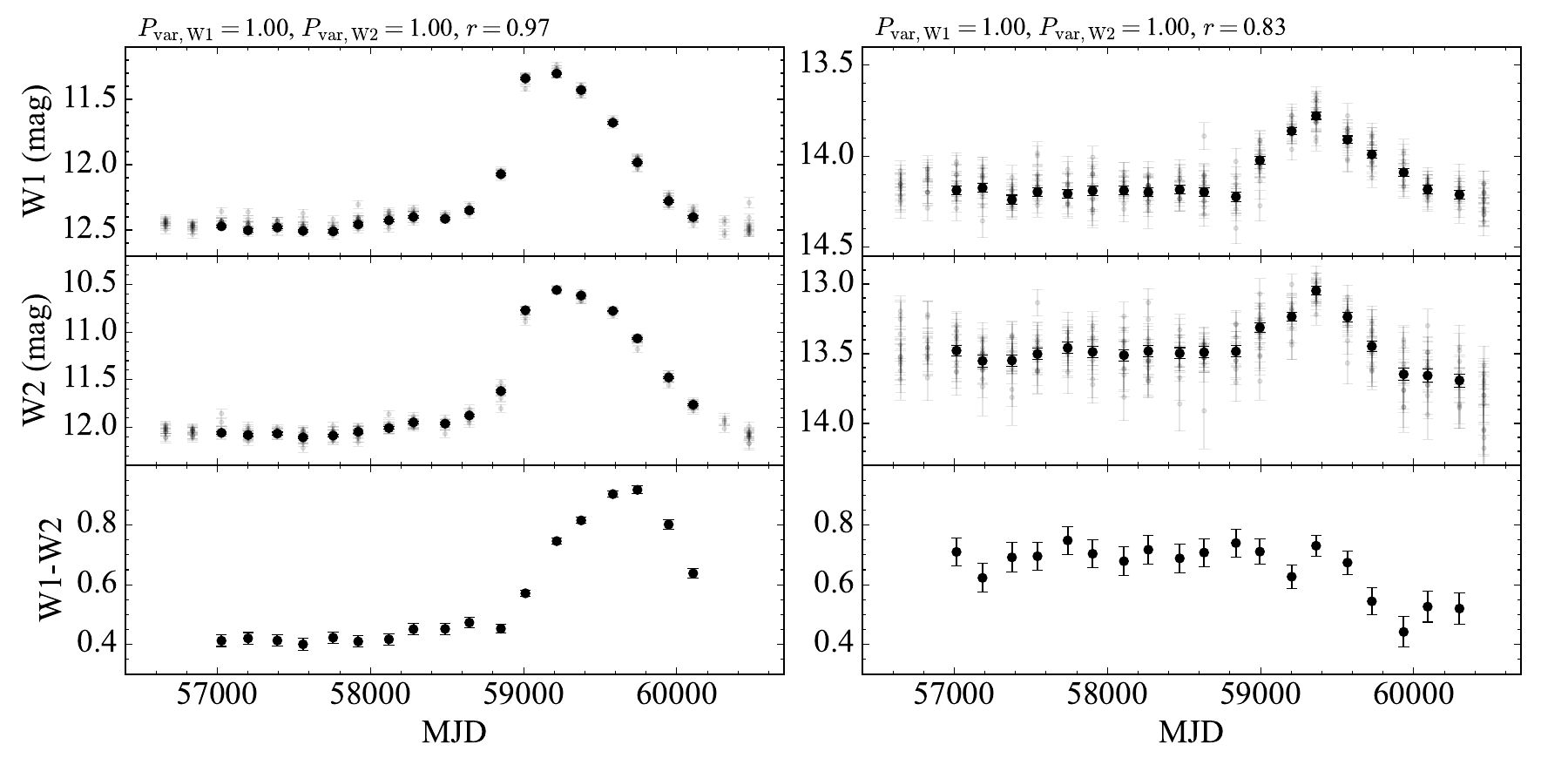}
\caption{Two examples of light curves showing the transient-like phenomena. The top and middle panels are the same as in Figure 1, while the bottom panels represent the W1$-$W2 color.  
}
\end{figure*}

\subsection{Caveat}
MIR variability can arise from sources other than AGNs, including rare events such as supernovae (SNe) and tidal disruption events (TDEs). Therefore, their contribution should be addressed properly. For TDEs, their low incident rate, estimated between $10^{-5}$ and $10^{-4}\ {\rm yr^{-1}}$ \cite[e.g.,][]{stone_2016}, makes them infrequent contaminants. Their rate is further suppressed for the black hole masses in our sample, as the TDE rate peaks around $M_{\rm BH} \sim 10^{6.5}M_\odot$ and declines at higher masses. Additionally, the dust obscuring TDEs is minimal, with a covering factor of only up to $1\%$ \cite[e.g.][]{jiang_2021}. Therefore, given the $\sim 10$-year observational span of NEOWISE, the likelihood of TDE contamination is considered negligible.

The incident rate for SNe is significantly higher than for TDEs, and unlike TDEs, SNe can occur in the galactic outskirts. Furthermore, SNe are categorized into various types, with differing energy outputs and rates among the subgroups \cite[e.g.,][]{li_2011}. Moreover, the dust formation mechanism and covering factor in SNe remain debated \cite[e.g.,][]{szalai_2019}. Given these combined factors, quantifying the potential for SNe contamination is difficult. For this reason, we visually inspect the observed light curves of 1000 randomly selected SF galaxies that exhibit MIR variability according to the Group 2 criteria. We select SF galaxies because they are primarily affected by SNe, whose rate is known to be proportional to SF rates \cite[e.g.,][]{mannucci_2005}. Of these, we identify 11 transient-like light curves (i.e., a monotonic decrease of brightness occasionally followed by a sudden increase). Examples of such light curves are shown in Figure 12. This result suggests that the effect of SNe is also minimal.   

\section{CONCLUSION}
To investigate an AGN-selection technique based on MIR variability, we use a sample of low-redshift galaxies from the MPA–JHU catalog, including both active and inactive systems. We cross-match this sample with NEOWISE and extract nearly a decade of MIR time-series photometry, obtained at a typical cadence of $\sim6$ months. After performing a careful calibration to remove spurious measurements, we construct high-quality MIR light curves for all matched sources. Using this dataset, we identify AGN candidates based on three variability-diagnostic parameters: (1) the likelihood of deviation from a non-variable light curve, $P_{\rm var}$; (2) the correlation coefficient between the W1 and W2 multi-epoch light curves, $r$; and (3) the fraction of epochs exhibiting the red W1–W2 color characteristic of known AGNs, $P_{\rm AGN}$. From this analysis, we draw the following conclusions:
\begin{itemize}

\item{From a fiducial test using optically selected AGNs, we find that AGNs can be reliably identified by applying two criteria: (1) $P_{\rm var, W1} > 0.99$ and $P_{\rm var, W2} > 0.99$ and (2) $r > 0.75$. With these selection requirements, we successfully identify AGNs for $\sim28\%$ of the optically selected AGN sample.}

\item{Applying conventional MIR color–based selection criteria reduces the AGN recovery fraction from $\sim28\%$ to $\sim9\%$, indicating that MIR color selection alone is inefficient for identifying AGNs whose MIR emission is strongly contaminated by stellar light or SF regions.}

\item{By applying the same criteria, we identify AGN candidates for $\sim12\%$, $\sim 3\%$, and $\sim 1\%$ of composites, SF galaxies, and normal galaxies, respectively. This suggests that AGNs can be missed when using only optical emission lines for classification.}

\item{Comparisons with other AGN diagnostics show that our MIR variability–based selection is broadly consistent with alternative methods, although non-negligible discrepancies remain among different AGN indicators. These results suggest that our approach is most effective for luminous AGNs and for identifying AGN candidates in star formation–dominated systems, while its overall completeness remains relatively low.}

\item{By comparing the AGN fraction ($f_{\rm var}$) derived from our MIR-based method with the offset from the SFMS ($\Delta {\rm SFMS}$) in the $M_*-SFR$ plane, we find that $f_{\rm var}$ increases with increasing of $\Delta {\rm SFMS}$. This trend suggests that AGN activity is closely connected to SF, providing evidence for the coevolution of SMBHs and their host galaxies.}

\item{For classical LINERs, the MIR variability-based AGN fraction is significantly lower than that of classical AGNs. This result is consistent with the fact that LINERs host predominantly low-luminosity AGNs or may lack a substantial dusty torus as a consequence of their low accretion luminosities.}

\item{The MIR variability can also be produced by other transient phenomena, such as SNe or TDEs. However, their occurrence rates are expected to be negligible and therefore have little impact on the conclusions of this study.}
\end{itemize}

\begin{acknowledgments}
We are grateful to the anonymous referee for valuable comments and suggestions, which greatly improved our manuscript.
LCH was supported by the National Science Foundation of China (12233001) and the China Manned Space Program (CMS-CSST-2025-A09). This work was supported by the National Research Foundation of Korea (NRF) grant funded by the Korean government (MSIT) (Nos. RS-2024-00347548 and RS-2025-16066624) and the Yonsei University Research Fund of 2025 (2025-22-0402).  
\end{acknowledgments}

\facilities{WISE, NEOWISE}
\software{Astropy \citep{2013A&A...558A..33A,2018AJ....156..123A,2022ApJ...935..167A}, SciPy \citep{2020SciPy}, NumPy \citep{2020numpy}, Matplotlib \citep{2007matplotlib}, scikit-learn \citep{2011scikit-learn}, pandas \citep{2020pandas}, NetworkX \citep{NetworkX}}

\bibliography{sample701}{}

\begin{thebibliography}{}
\expandafter\ifx\csname natexlab\endcsname\relax\def\natexlab#1{#1}\fi
\providecommand{\url}[1]{\href{#1}{#1}}
\providecommand{\dodoi}[1]{doi:~\href{http://doi.org/#1}{\nolinkurl{#1}}}
\providecommand{\doeprint}[1]{\href{http://ascl.net/#1}{\nolinkurl{http://ascl.net/#1}}}
\providecommand{\doarXiv}[1]{\href{https://arxiv.org/abs/#1}{\nolinkurl{https://arxiv.org/abs/#1}}}

\bibitem[{K. {Abazajian} {et~al.}(2003){Abazajian}, {Adelman-McCarthy}, {Ag{\"u}eros}, {Allam}, {Anderson}, {Annis}, {Bahcall}, {Baldry}, {Bastian}, {Berlind}, {Bernardi}, {Blanton}, {Blythe}, {Bochanski}, {Boroski}, {Brewington}, {Briggs}, {Brinkmann}, {Brunner}, {Budav{\'a}ri}, {Carey}, {Carr}, {Castander}, {Chiu}, {Collinge}, {Connolly}, {Covey}, {Csabai}, {Dalcanton}, {Dodelson}, {Doi}, {Dong}, {Eisenstein}, {Evans}, {Fan}, {Feldman}, {Finkbeiner}, {Friedman}, {Frieman}, {Fukugita}, {Gal}, {Gillespie}, {Glazebrook}, {Gonzalez}, {Gray}, {Grebel}, {Grodnicki}, {Gunn}, {Gurbani}, {Hall}, {Hao}, {Harbeck}, {Harris}, {Harris}, {Harvanek}, {Hawley}, {Heckman}, {Helmboldt}, {Hendry}, {Hennessy}, {Hindsley}, {Hogg}, {Holmgren}, {Holtzman}, {Homer}, {Hui}, {Ichikawa}, {Ichikawa}, {Inkmann}, {Ivezi{\'c}}, {Jester}, {Johnston}, {Jordan}, {Jordan}, {Jorgensen}, {Juri{\'c}}, {Kauffmann}, {Kent}, {Kleinman}, {Knapp}, {Kniazev}, {Kron}, {Krzesi{\'n}ski}, {Kunszt}, {Kuropatkin}, {Lamb}, {Lampeitl}, {Laubscher}, {Lee},
  {Leger}, {Li}, {Lidz}, {Lin}, {Loh}, {Long}, {Loveday}, {Lupton}, {Malik}, {Margon}, {McGehee}, {McKay}, {Meiksin}, {Miknaitis}, {Moorthy}, {Munn}, {Murphy}, {Nakajima}, {Narayanan}, {Nash}, {Neilsen}, {Newberg}, {Newman}, {Nichol}, {Nicinski}, {Nieto-Santisteban}, {Nitta}, {Odenkirchen}, {Okamura}, {Ostriker}, {Owen}, {Padmanabhan}, {Peoples}, {Pier}, {Pindor}, {Pope}, {Quinn}, {Rafikov}, {Raymond}, {Richards}, {Richmond}, {Rix}, {Rockosi}, {Schaye}, {Schlegel}, {Schneider}, {Schroeder}, {Scranton}, {Sekiguchi}, {Seljak}, {Sergey}, {Sesar}, {Sheldon}, {Shimasaku}, {Siegmund}, {Silvestri}, {Sinisgalli}, {Sirko}, {Smith}, {Smol{\v{c}}i{\'c}}, {Snedden}, {Stebbins}, {Steinhardt}, {Stinson}, {Stoughton}, {Strateva}, {Strauss}, {SubbaRao}, {Szalay}, {Szapudi}, {Szkody}, {Tasca}, {Tegmark}, {Thakar}, {Tremonti}, {Tucker}, {Uomoto}, {Vanden Berk}, {Vandenberg}, {Vogeley}, {Voges}, {Vogt}, {Walkowicz}, {Weinberg}, {West}, {White}, {Wilhite}, {Willman}, {Xu}, {Yanny}, {Yarger}, {Yasuda}, {Yip}, {Yocum}, {York},
  {Zakamska}, {Zehavi}, {Zheng}, {Zibetti}, \& {Zucker}}]{sdssdr1_2003}
{Abazajian}, K., {Adelman-McCarthy}, J.~K., {Ag{\"u}eros}, M.~A., {et~al.} 2003, \bibinfo{title}{{The First Data Release of the Sloan Digital Sky Survey},} \aj, 126, 2081, \dodoi{10.1086/378165}

\bibitem[{C.~J. {Agostino} \& S. {Salim}(2019){Agostino} \& {Salim}}]{agostino_2019}
{Agostino}, C.~J., \& {Salim}, S. 2019, \bibinfo{title}{{Crossing the Line: Active Galactic Nuclei in the Star-forming Region of the BPT Diagram},} \apj, 876, 12, \dodoi{10.3847/1538-4357/ab1094}

\bibitem[{H. {Aihara} {et~al.}(2011){Aihara}, {Allende Prieto}, {An}, {Anderson}, {Aubourg}, {Balbinot}, {Beers}, {Berlind}, {Bickerton}, {Bizyaev}, {Blanton}, {Bochanski}, {Bolton}, {Bovy}, {Brandt}, {Brinkmann}, {Brown}, {Brownstein}, {Busca}, {Campbell}, {Carr}, {Chen}, {Chiappini}, {Comparat}, {Connolly}, {Cortes}, {Croft}, {Cuesta}, {da Costa}, {Davenport}, {Dawson}, {Dhital}, {Ealet}, {Ebelke}, {Edmondson}, {Eisenstein}, {Escoffier}, {Esposito}, {Evans}, {Fan}, {Femen{\'\i}a Castell{\'a}}, {Font-Ribera}, {Frinchaboy}, {Ge}, {Gillespie}, {Gilmore}, {Gonz{\'a}lez Hern{\'a}ndez}, {Gott}, {Gould}, {Grebel}, {Gunn}, {Hamilton}, {Harding}, {Harris}, {Hawley}, {Hearty}, {Ho}, {Hogg}, {Holtzman}, {Honscheid}, {Inada}, {Ivans}, {Jiang}, {Johnson}, {Jordan}, {Jordan}, {Kazin}, {Kirkby}, {Klaene}, {Knapp}, {Kneib}, {Kochanek}, {Koesterke}, {Kollmeier}, {Kron}, {Lampeitl}, {Lang}, {Le Goff}, {Lee}, {Lin}, {Long}, {Loomis}, {Lucatello}, {Lundgren}, {Lupton}, {Ma}, {MacDonald}, {Mahadevan}, {Maia}, {Makler},
  {Malanushenko}, {Malanushenko}, {Mandelbaum}, {Maraston}, {Margala}, {Masters}, {McBride}, {McGehee}, {McGreer}, {M{\'e}nard}, {Miralda-Escud{\'e}}, {Morrison}, {Mullally}, {Muna}, {Munn}, {Murayama}, {Myers}, {Naugle}, {Neto}, {Nguyen}, {Nichol}, {O'Connell}, {Ogando}, {Olmstead}, {Oravetz}, {Padmanabhan}, {Palanque-Delabrouille}, {Pan}, {Pandey}, {P{\^a}ris}, {Percival}, {Petitjean}, {Pfaffenberger}, {Pforr}, {Phleps}, {Pichon}, {Pieri}, {Prada}, {Price-Whelan}, {Raddick}, {Ramos}, {Reyl{\'e}}, {Rich}, {Richards}, {Rix}, {Robin}, {Rocha-Pinto}, {Rockosi}, {Roe}, {Rollinde}, {Ross}, {Ross}, {Rossetto}, {S{\'a}nchez}, {Sayres}, {Schlegel}, {Schlesinger}, {Schmidt}, {Schneider}, {Sheldon}, {Shu}, {Simmerer}, {Simmons}, {Sivarani}, {Snedden}, {Sobeck}, {Steinmetz}, {Strauss}, {Szalay}, {Tanaka}, {Thakar}, {Thomas}, {Tinker}, {Tofflemire}, {Tojeiro}, {Tremonti}, {Vandenberg}, {Vargas Maga{\~n}a}, {Verde}, {Vogt}, {Wake}, {Wang}, {Weaver}, {Weinberg}, {White}, {White}, {Yanny}, {Yasuda}, {Yeche}, \&
  {Zehavi}}]{aihara_2011}
{Aihara}, H., {Allende Prieto}, C., {An}, D., {et~al.} 2011, \bibinfo{title}{{The Eighth Data Release of the Sloan Digital Sky Survey: First Data from SDSS-III},} \apjs, 193, 29, \dodoi{10.1088/0067-0049/193/2/29}

\bibitem[{R. {Antonucci}(1993){Antonucci}}]{antonucci_1993}
{Antonucci}, R. 1993, \bibinfo{title}{{Unified models for active galactic nuclei and quasars.},} \araa, 31, 473, \dodoi{10.1146/annurev.aa.31.090193.002353}

\bibitem[{A.~S. {Aradhey} {et~al.}(2025){Aradhey}, {Constantin}, {Vogeley}, \& {Douglass}}]{2025ApJ...991...52A}
{Aradhey}, A.~S., {Constantin}, A., {Vogeley}, M.~S., \& {Douglass}, K.~A. 2025, \bibinfo{title}{{Quantifying the Active Galactic Nucleus Fraction in Cosmic Voids via Mid-infrared Variability},} \apj, 991, 52, \dodoi{10.3847/1538-4357/adeca1}

\bibitem[{P. {Ar{\'e}valo} {et~al.}(2026){Ar{\'e}valo}, {S{\'a}nchez-S{\'a}ez}, {Sotomayor}, {Lira}, {Bauer}, \& {R{\'\i}os}}]{arevalo_2026}
{Ar{\'e}valo}, P., {S{\'a}nchez-S{\'a}ez}, P., {Sotomayor}, B., {et~al.} 2026, \bibinfo{title}{{Unlocking AGN variability with custom ZTF photometry for high-fidelity light curves and robust selection},} \aap, 705, A247, \dodoi{10.1051/0004-6361/202556258}

\bibitem[{R.~J. {Assef} {et~al.}(2018){Assef}, {Stern}, {Noirot}, {Jun}, {Cutri}, \& {Eisenhardt}}]{assef_2018}
{Assef}, R.~J., {Stern}, D., {Noirot}, G., {et~al.} 2018, \bibinfo{title}{{The WISE AGN Catalog},} \apjs, 234, 23, \dodoi{10.3847/1538-4365/aaa00a}

\bibitem[{ {Astropy Collaboration} {et~al.}(2013){Astropy Collaboration}, {Robitaille}, {Tollerud}, {Greenfield}, {Droettboom}, {Bray}, {Aldcroft}, {Davis}, {Ginsburg}, {Price-Whelan}, {Kerzendorf}, {Conley}, {Crighton}, {Barbary}, {Muna}, {Ferguson}, {Grollier}, {Parikh}, {Nair}, {Unther}, {Deil}, {Woillez}, {Conseil}, {Kramer}, {Turner}, {Singer}, {Fox}, {Weaver}, {Zabalza}, {Edwards}, {Azalee Bostroem}, {Burke}, {Casey}, {Crawford}, {Dencheva}, {Ely}, {Jenness}, {Labrie}, {Lim}, {Pierfederici}, {Pontzen}, {Ptak}, {Refsdal}, {Servillat}, \& {Streicher}}]{2013A&A...558A..33A}
{Astropy Collaboration}, {Robitaille}, T.~P., {Tollerud}, E.~J., {et~al.} 2013, \bibinfo{title}{{Astropy: A community Python package for astronomy},} \aap, 558, A33, \dodoi{10.1051/0004-6361/201322068}

\bibitem[{ {Astropy Collaboration} {et~al.}(2018){Astropy Collaboration}, {Price-Whelan}, {Sip{\H{o}}cz}, {G{\"u}nther}, {Lim}, {Crawford}, {Conseil}, {Shupe}, {Craig}, {Dencheva}, {Ginsburg}, {VanderPlas}, {Bradley}, {P{\'e}rez-Su{\'a}rez}, {de Val-Borro}, {Aldcroft}, {Cruz}, {Robitaille}, {Tollerud}, {Ardelean}, {Babej}, {Bach}, {Bachetti}, {Bakanov}, {Bamford}, {Barentsen}, {Barmby}, {Baumbach}, {Berry}, {Biscani}, {Boquien}, {Bostroem}, {Bouma}, {Brammer}, {Bray}, {Breytenbach}, {Buddelmeijer}, {Burke}, {Calderone}, {Cano Rodr{\'\i}guez}, {Cara}, {Cardoso}, {Cheedella}, {Copin}, {Corrales}, {Crichton}, {D'Avella}, {Deil}, {Depagne}, {Dietrich}, {Donath}, {Droettboom}, {Earl}, {Erben}, {Fabbro}, {Ferreira}, {Finethy}, {Fox}, {Garrison}, {Gibbons}, {Goldstein}, {Gommers}, {Greco}, {Greenfield}, {Groener}, {Grollier}, {Hagen}, {Hirst}, {Homeier}, {Horton}, {Hosseinzadeh}, {Hu}, {Hunkeler}, {Ivezi{\'c}}, {Jain}, {Jenness}, {Kanarek}, {Kendrew}, {Kern}, {Kerzendorf}, {Khvalko}, {King}, {Kirkby}, {Kulkarni},
  {Kumar}, {Lee}, {Lenz}, {Littlefair}, {Ma}, {Macleod}, {Mastropietro}, {McCully}, {Montagnac}, {Morris}, {Mueller}, {Mumford}, {Muna}, {Murphy}, {Nelson}, {Nguyen}, {Ninan}, {N{\"o}the}, {Ogaz}, {Oh}, {Parejko}, {Parley}, {Pascual}, {Patil}, {Patil}, {Plunkett}, {Prochaska}, {Rastogi}, {Reddy Janga}, {Sabater}, {Sakurikar}, {Seifert}, {Sherbert}, {Sherwood-Taylor}, {Shih}, {Sick}, {Silbiger}, {Singanamalla}, {Singer}, {Sladen}, {Sooley}, {Sornarajah}, {Streicher}, {Teuben}, {Thomas}, {Tremblay}, {Turner}, {Terr{\'o}n}, {van Kerkwijk}, {de la Vega}, {Watkins}, {Weaver}, {Whitmore}, {Woillez}, {Zabalza}, \& {Astropy Contributors}}]{2018AJ....156..123A}
{Astropy Collaboration}, {Price-Whelan}, A.~M., {Sip{\H{o}}cz}, B.~M., {et~al.} 2018, \bibinfo{title}{{The Astropy Project: Building an Open-science Project and Status of the v2.0 Core Package},} \aj, 156, 123, \dodoi{10.3847/1538-3881/aabc4f}

\bibitem[{ {Astropy Collaboration} {et~al.}(2022){Astropy Collaboration}, {Price-Whelan}, {Lim}, {Earl}, {Starkman}, {Bradley}, {Shupe}, {Patil}, {Corrales}, {Brasseur}, {N{\"o}the}, {Donath}, {Tollerud}, {Morris}, {Ginsburg}, {Vaher}, {Weaver}, {Tocknell}, {Jamieson}, {van Kerkwijk}, {Robitaille}, {Merry}, {Bachetti}, {G{\"u}nther}, {Aldcroft}, {Alvarado-Montes}, {Archibald}, {B{\'o}di}, {Bapat}, {Barentsen}, {Baz{\'a}n}, {Biswas}, {Boquien}, {Burke}, {Cara}, {Cara}, {Conroy}, {Conseil}, {Craig}, {Cross}, {Cruz}, {D'Eugenio}, {Dencheva}, {Devillepoix}, {Dietrich}, {Eigenbrot}, {Erben}, {Ferreira}, {Foreman-Mackey}, {Fox}, {Freij}, {Garg}, {Geda}, {Glattly}, {Gondhalekar}, {Gordon}, {Grant}, {Greenfield}, {Groener}, {Guest}, {Gurovich}, {Handberg}, {Hart}, {Hatfield-Dodds}, {Homeier}, {Hosseinzadeh}, {Jenness}, {Jones}, {Joseph}, {Kalmbach}, {Karamehmetoglu}, {Ka{\l}uszy{\'n}ski}, {Kelley}, {Kern}, {Kerzendorf}, {Koch}, {Kulumani}, {Lee}, {Ly}, {Ma}, {MacBride}, {Maljaars}, {Muna}, {Murphy}, {Norman},
  {O'Steen}, {Oman}, {Pacifici}, {Pascual}, {Pascual-Granado}, {Patil}, {Perren}, {Pickering}, {Rastogi}, {Roulston}, {Ryan}, {Rykoff}, {Sabater}, {Sakurikar}, {Salgado}, {Sanghi}, {Saunders}, {Savchenko}, {Schwardt}, {Seifert-Eckert}, {Shih}, {Jain}, {Shukla}, {Sick}, {Simpson}, {Singanamalla}, {Singer}, {Singhal}, {Sinha}, {Sip{\H{o}}cz}, {Spitler}, {Stansby}, {Streicher}, {{\v{S}}umak}, {Swinbank}, {Taranu}, {Tewary}, {Tremblay}, {de Val-Borro}, {Van Kooten}, {Vasovi{\'c}}, {Verma}, {de Miranda Cardoso}, {Williams}, {Wilson}, {Winkel}, {Wood-Vasey}, {Xue}, {Yoachim}, {Zhang}, {Zonca}, \& {Astropy Project Contributors}}]{2022ApJ...935..167A}
{Astropy Collaboration}, {Price-Whelan}, A.~M., {Lim}, P.~L., {et~al.} 2022, \bibinfo{title}{{The Astropy Project: Sustaining and Growing a Community-oriented Open-source Project and the Latest Major Release (v5.0) of the Core Package},} \apj, 935, 167, \dodoi{10.3847/1538-4357/ac7c74}

\bibitem[{J.~A. {Baldwin} {et~al.}(1981){Baldwin}, {Phillips}, \& {Terlevich}}]{baldwin_1981}
{Baldwin}, J.~A., {Phillips}, M.~M., \& {Terlevich}, R. 1981, \bibinfo{title}{{Classification parameters for the emission-line spectra of extragalactic objects.},} \pasp, 93, 5, \dodoi{10.1086/130766}

\bibitem[{E.~C. {Bellm} {et~al.}(2019){Bellm}, {Kulkarni}, {Graham}, {Dekany}, {Smith}, {Riddle}, {Masci}, {Helou}, {Prince}, {Adams}, {Barbarino}, {Barlow}, {Bauer}, {Beck}, {Belicki}, {Biswas}, {Blagorodnova}, {Bodewits}, {Bolin}, {Brinnel}, {Brooke}, {Bue}, {Bulla}, {Burruss}, {Cenko}, {Chang}, {Connolly}, {Coughlin}, {Cromer}, {Cunningham}, {De}, {Delacroix}, {Desai}, {Duev}, {Eadie}, {Farnham}, {Feeney}, {Feindt}, {Flynn}, {Franckowiak}, {Frederick}, {Fremling}, {Gal-Yam}, {Gezari}, {Giomi}, {Goldstein}, {Golkhou}, {Goobar}, {Groom}, {Hacopians}, {Hale}, {Henning}, {Ho}, {Hover}, {Howell}, {Hung}, {Huppenkothen}, {Imel}, {Ip}, {Ivezi{\'c}}, {Jackson}, {Jones}, {Juric}, {Kasliwal}, {Kaspi}, {Kaye}, {Kelley}, {Kowalski}, {Kramer}, {Kupfer}, {Landry}, {Laher}, {Lee}, {Lin}, {Lin}, {Lunnan}, {Giomi}, {Mahabal}, {Mao}, {Miller}, {Monkewitz}, {Murphy}, {Ngeow}, {Nordin}, {Nugent}, {Ofek}, {Patterson}, {Penprase}, {Porter}, {Rauch}, {Rebbapragada}, {Reiley}, {Rigault}, {Rodriguez}, {van Roestel}, {Rusholme},
  {van Santen}, {Schulze}, {Shupe}, {Singer}, {Soumagnac}, {Stein}, {Surace}, {Sollerman}, {Szkody}, {Taddia}, {Terek}, {Van Sistine}, {van Velzen}, {Vestrand}, {Walters}, {Ward}, {Ye}, {Yu}, {Yan}, \& {Zolkower}}]{bellm_2019}
{Bellm}, E.~C., {Kulkarni}, S.~R., {Graham}, M.~J., {et~al.} 2019, \bibinfo{title}{{The Zwicky Transient Facility: System Overview, Performance, and First Results},} \pasp, 131, 018002, \dodoi{10.1088/1538-3873/aaecbe}

\bibitem[{P.~N. {Best} \& T.~M. {Heckman}(2012){Best} \& {Heckman}}]{best_2012}
{Best}, P.~N., \& {Heckman}, T.~M. 2012, \bibinfo{title}{{On the fundamental dichotomy in the local radio-AGN population: accretion, evolution and host galaxy properties},} \mnras, 421, 1569, \dodoi{10.1111/j.1365-2966.2012.20414.x}

\bibitem[{J.~J. {Bock} {et~al.}(2026){Bock}, {Aboobaker}, {Adamo}, {Akeson}, {Alred}, {Alibay}, {Ashby}, {Bach}, {Bleem}, {Bolton}, {Braun}, {Bruton}, {Bryan}, {Chang}, {Chen}, {Cheng}, {Cheshire}, {Chiang}, {Choppin de Janvry}, {Condon}, {Cook}, {Cooray}, {Crill}, {Cukierman}, {Dore}, {Dowell}, {Dubois-Felsmann}, {Eifler}, {Everett}, {Fabinsky}, {Faisst}, {Fanson}, {Farrington}, {Fatahi}, {Fazar}, {Feder}, {Frater}, {Grasshorn Gebhardt}, {Giri}, {Goldina}, {Gorjian}, {Habib}, {Hart}, {Heinrich}, {Hora}, {Huai}, {Hui}, {Jo}, {Jeong}, {Kang}, {Kang}, {Kecman}, {Kim}, {Kim}, {Kim}, {Kim}, {Kim}, {Kirkpatrick}, {Kobayashi}, {Korngut}, {Krause}, {Lee}, {Lee}, {Lee}, {Lee}, {Lisse}, {Mariani}, {Masters}, {Mauskopf}, {Melnick}, {Minasyan}, {Mirocha}, {Miyasaka}, {Moore}, {Moore}, {Murgia}, {Naylor}, {Nelson}, {Nguyen}, {Nguyen}, {Noh}, {Padin}, {Paladini}, {Park}, {Penanen}, {Putnam}, {Pyo}, {Ramachandra}, {Ramanathan}, {Rustamkulov}, {Reiley}, {Rice}, {Rocca}, {Seok}, {Smith}, {Stober}, {Susca}, {Teplitz},
  {Thelen}, {Tolls}, {Torrini}, {Trangsrud}, {Unwin}, {Velicheti}, {Wang}, {Wen}, {-Werner}, {Williams}, {Williamson}, {Wincentsen}, {Windhorst}, {Yang}, {Yang}, \& {Zemcov}}]{bock_2025}
{Bock}, J.~J., {Aboobaker}, A.~M., {Adamo}, J., {et~al.} 2026, \bibinfo{title}{{The SPHEREx Satellite Mission},} arXiv e-prints, arXiv:2511.02985, \dodoi{10.48550/arXiv.2511.02985}

\bibitem[{J. {Brinchmann} {et~al.}(2004){Brinchmann}, {Charlot}, {White}, {Tremonti}, {Kauffmann}, {Heckman}, \& {Brinkmann}}]{2004MNRAS.351.1151B}
{Brinchmann}, J., {Charlot}, S., {White}, S.~D.~M., {et~al.} 2004, \bibinfo{title}{{The physical properties of star-forming galaxies in the low-redshift Universe},} \mnras, 351, 1151, \dodoi{10.1111/j.1365-2966.2004.07881.x}

\bibitem[{W. {Byun} {et~al.}(2023){Byun}, {Kim}, {Sheen}, {Lee}, {Ho}, {Ko}, {Seon}, {Shim}, {Kim}, {Kim}, {Lee}, {Jeong}, {Woo}, {Jeong}, {Park}, {Kim}, {Lee}, {Cha}, {Song}, {Son}, \& {Yang}}]{byun_2023}
{Byun}, W., {Kim}, M., {Sheen}, Y.-K., {et~al.} 2023, \bibinfo{title}{{Photometric Selection of Unobscured QSOs at the Ecliptic Poles: KMTNet in the South Field and Pan-STARRS in the North Field},} \apjs, 268, 57, \dodoi{10.3847/1538-4365/acebe4}

\bibitem[{Y. {Choi} {et~al.}(2014){Choi}, {Gibson}, {Becker}, {Ivezi{\'c}}, {Connolly}, {MacLeod}, {Ruan}, \& {Anderson}}]{choi_2014}
{Choi}, Y., {Gibson}, R.~R., {Becker}, A.~C., {et~al.} 2014, \bibinfo{title}{{Variability-based Active Galactic Nucleus Selection Using Image Subtraction in the SDSS and LSST Era},} \apj, 782, 37, \dodoi{10.1088/0004-637X/782/1/37}

\bibitem[{L. {Ciesla} {et~al.}(2015){Ciesla}, {Charmandaris}, {Georgakakis}, {Bernhard}, {Mitchell}, {Buat}, {Elbaz}, {LeFloc'h}, {Lacey}, {Magdis}, \& {Xilouris}}]{ciesla_2015}
{Ciesla}, L., {Charmandaris}, V., {Georgakakis}, A., {et~al.} 2015, \bibinfo{title}{{Constraining the properties of AGN host galaxies with spectral energy distribution modelling},} \aap, 576, A10, \dodoi{10.1051/0004-6361/201425252}

\bibitem[{A.~M. {Diamond-Stanic} \& G.~H. {Rieke}(2012){Diamond-Stanic} \& {Rieke}}]{diamond_2012}
{Diamond-Stanic}, A.~M., \& {Rieke}, G.~H. 2012, \bibinfo{title}{{The Relationship between Black Hole Growth and Star Formation in Seyfert Galaxies},} \apj, 746, 168, \dodoi{10.1088/0004-637X/746/2/168}

\bibitem[{J.~L. {Donley} {et~al.}(2012){Donley}, {Koekemoer}, {Brusa}, {Capak}, {Cardamone}, {Civano}, {Ilbert}, {Impey}, {Kartaltepe}, {Miyaji}, {Salvato}, {Sanders}, {Trump}, \& {Zamorani}}]{donley_2012}
{Donley}, J.~L., {Koekemoer}, A.~M., {Brusa}, M., {et~al.} 2012, \bibinfo{title}{{Identifying Luminous Active Galactic Nuclei in Deep Surveys: Revised IRAC Selection Criteria},} \apj, 748, 142, \dodoi{10.1088/0004-637X/748/2/142}

\bibitem[{M. {Elitzur} \& L.~C. {Ho}(2009){Elitzur} \& {Ho}}]{elitzur_2009}
{Elitzur}, M., \& {Ho}, L.~C. 2009, \bibinfo{title}{{On the Disappearance of the Broad-Line Region in Low-Luminosity Active Galactic Nuclei},} \apjl, 701, L91, \dodoi{10.1088/0004-637X/701/2/L91}

\bibitem[{M. {Elitzur} \& I. {Shlosman}(2006){Elitzur} \& {Shlosman}}]{elitzur_2006}
{Elitzur}, M., \& {Shlosman}, I. 2006, \bibinfo{title}{{The AGN-obscuring Torus: The End of the ``Doughnut'' Paradigm?},} \apjl, 648, L101, \dodoi{10.1086/508158}

\bibitem[{Y.~A. {Gordon} {et~al.}(2020){Gordon}, {Boyce}, {O'Dea}, {Rudnick}, {Andernach}, {Vantyghem}, {Baum}, {Bui}, \& {Dionyssiou}}]{gordon_2020}
{Gordon}, Y.~A., {Boyce}, M.~M., {O'Dea}, C.~P., {et~al.} 2020, \bibinfo{title}{{A Catalog of Very Large Array Sky Survey Epoch 1 Quick Look Components, Sources, and Host Identifications},} Research Notes of the American Astronomical Society, 4, 175, \dodoi{10.3847/2515-5172/abbe23}

\bibitem[{J.~E. {Greene} {et~al.}(2020){Greene}, {Strader}, \& {Ho}}]{greene_2020}
{Greene}, J.~E., {Strader}, J., \& {Ho}, L.~C. 2020, \bibinfo{title}{{Intermediate-Mass Black Holes},} \araa, 58, 257, \dodoi{10.1146/annurev-astro-032620-021835}

\bibitem[{A.~A. Hagberg {et~al.}(2008)Hagberg, Schult, \& Swart}]{NetworkX}
Hagberg, A.~A., Schult, D.~A., \& Swart, P.~J. 2008, \bibinfo{title}{Exploring Network Structure, Dynamics, and Function using NetworkX,} in Proceedings of the 7th Python in Science Conference, ed. G.~Varoquaux, T.~Vaught, \& J.~Millman, Pasadena, CA USA, 11 -- 15

\bibitem[{C.~R. Harris {et~al.}(2020)Harris, Millman, van~der Walt, Gommers, Virtanen, Cournapeau, Wieser, Taylor, Berg, Smith, Kern, Picus, Hoyer, van Kerkwijk, Brett, Haldane, del R{\'{i}}o, Wiebe, Peterson, G{\'{e}}rard-Marchant, Sheppard, Reddy, Weckesser, Abbasi, Gohlke, \& Oliphant}]{2020numpy}
Harris, C.~R., Millman, K.~J., van~der Walt, S.~J., {et~al.} 2020, \bibinfo{title}{Array programming with {NumPy},} Nature, 585, 357, \dodoi{10.1038/s41586-020-2649-2}

\bibitem[{T.~M. {Heckman} {et~al.}(2005){Heckman}, {Ptak}, {Hornschemeier}, \& {Kauffmann}}]{heckman_2005}
{Heckman}, T.~M., {Ptak}, A., {Hornschemeier}, A., \& {Kauffmann}, G. 2005, \bibinfo{title}{{The Relationship of Hard X-Ray and Optical Line Emission in Low-Redshift Active Galactic Nuclei},} \apj, 634, 161, \dodoi{10.1086/491665}

\bibitem[{R.~C. {Hickox} \& D.~M. {Alexander}(2018){Hickox} \& {Alexander}}]{hickox_2018}
{Hickox}, R.~C., \& {Alexander}, D.~M. 2018, \bibinfo{title}{{Obscured Active Galactic Nuclei},} \araa, 56, 625, \dodoi{10.1146/annurev-astro-081817-051803}

\bibitem[{L.~C. {Ho}(1999){Ho}}]{ho_1999}
{Ho}, L.~C. 1999, \bibinfo{title}{{The Spectral Energy Distributions of Low-Luminosity Active Galactic Nuclei},} \apj, 516, 672, \dodoi{10.1086/307137}

\bibitem[{L.~C. {Ho}(2002){Ho}}]{ho_2002}
{Ho}, L.~C. 2002, \bibinfo{title}{{On the Relationship between Radio Emission and Black Hole Mass in Galactic Nuclei},} \apj, 564, 120, \dodoi{10.1086/324399}

\bibitem[{L.~C. {Ho}(2005){Ho}}]{ho_2005}
{Ho}, L.~C. 2005, \bibinfo{title}{{[O II] Emission in Quasar Host Galaxies: Evidence for a Suppressed Star Formation Efficiency},} \apj, 629, 680, \dodoi{10.1086/431643}

\bibitem[{L.~C. {Ho}(2008){Ho}}]{ho_2008}
{Ho}, L.~C. 2008, \bibinfo{title}{{Nuclear activity in nearby galaxies.},} \araa, 46, 475, \dodoi{10.1146/annurev.astro.45.051806.110546}

\bibitem[{L.~C. {Ho}(2009){Ho}}]{ho_2009}
{Ho}, L.~C. 2009, \bibinfo{title}{{Radiatively Inefficient Accretion in Nearby Galaxies},} \apj, 699, 626, \dodoi{10.1088/0004-637X/699/1/626}

\bibitem[{L.~C. {Ho} {et~al.}(1995){Ho}, {Filippenko}, \& {Sargent}}]{ho_1995}
{Ho}, L.~C., {Filippenko}, A.~V., \& {Sargent}, W.~L. 1995, \bibinfo{title}{{A Search for ``Dwarf'' Seyfert Nuclei. II. an Optical Spectral Atlas of the Nuclei of Nearby Galaxies},} \apjs, 98, 477, \dodoi{10.1086/192170}

\bibitem[{L.~C. {Ho} {et~al.}(1993){Ho}, {Filippenko}, \& {Sargent}}]{ho_1993}
{Ho}, L.~C., {Filippenko}, A.~V., \& {Sargent}, W. L.~W. 1993, \bibinfo{title}{{A Reevaluation of the Excitation Mechanism of LINERs},} \apj, 417, 63, \dodoi{10.1086/173291}

\bibitem[{L.~C. {Ho} {et~al.}(1997{\natexlab{a}}){Ho}, {Filippenko}, \& {Sargent}}]{ho_1997}
{Ho}, L.~C., {Filippenko}, A.~V., \& {Sargent}, W. L.~W. 1997{\natexlab{a}}, \bibinfo{title}{{A Search for ``Dwarf'' Seyfert Nuclei. V. Demographics of Nuclear Activity in Nearby Galaxies},} \apj, 487, 568, \dodoi{10.1086/304638}

\bibitem[{L.~C. {Ho} {et~al.}(1997{\natexlab{b}}){Ho}, {Filippenko}, \& {Sargent}}]{ho_1997b}
{Ho}, L.~C., {Filippenko}, A.~V., \& {Sargent}, W. L.~W. 1997{\natexlab{b}}, \bibinfo{title}{{A Search for ``Dwarf'' Seyfert Nuclei. III. Spectroscopic Parameters and Properties of the Host Galaxies},} \apjs, 112, 315, \dodoi{10.1086/313041}

\bibitem[{L.~C. {Ho} {et~al.}(2012){Ho}, {Kim}, \& {Terashima}}]{ho_2012}
{Ho}, L.~C., {Kim}, M., \& {Terashima}, Y. 2012, \bibinfo{title}{{The Low-mass, Highly Accreting Black Hole Associated with the Active Galactic Nucleus 2XMM J123103.2+110648},} \apjl, 759, L16, \dodoi{10.1088/2041-8205/759/1/L16}

\bibitem[{L.~C. {Ho} \& C.~Y. {Peng}(2001){Ho} \& {Peng}}]{ho_2001}
{Ho}, L.~C., \& {Peng}, C.~Y. 2001, \bibinfo{title}{{Nuclear Luminosities and Radio Loudness of Seyfert Nuclei},} \apj, 555, 650, \dodoi{10.1086/321524}

\bibitem[{J.~D. Hunter(2007)Hunter}]{2007matplotlib}
Hunter, J.~D. 2007, \bibinfo{title}{Matplotlib: A 2D graphics environment,} Computing in Science \& Engineering, 9, 90, \dodoi{10.1109/MCSE.2007.55}

\bibitem[{M. {Imanishi}(2009){Imanishi}}]{imanishi_2009}
{Imanishi}, M. 2009, \bibinfo{title}{{Luminous Buried Active Galactic Nuclei as a Function of Galaxy Infrared Luminosity Revealed through Spitzer Low-resolution Infrared Spectroscopy},} \apj, 694, 751, \dodoi{10.1088/0004-637X/694/2/751}

\bibitem[{{\v{Z}}. {Ivezi{\'c}} {et~al.}(2002){Ivezi{\'c}}, {Menou}, {Knapp}, {Strauss}, {Lupton}, {Vanden Berk}, {Richards}, {Tremonti}, {Weinstein}, {Anderson}, {Bahcall}, {Becker}, {Bernardi}, {Blanton}, {Eisenstein}, {Fan}, {Finkbeiner}, {Finlator}, {Frieman}, {Gunn}, {Hall}, {Kim}, {Kinkhabwala}, {Narayanan}, {Rockosi}, {Schlegel}, {Schneider}, {Strateva}, {SubbaRao}, {Thakar}, {Voges}, {White}, {Yanny}, {Brinkmann}, {Doi}, {Fukugita}, {Hennessy}, {Munn}, {Nichol}, \& {York}}]{ivezic_2002}
{Ivezi{\'c}}, {\v{Z}}., {Menou}, K., {Knapp}, G.~R., {et~al.} 2002, \bibinfo{title}{{Optical and Radio Properties of Extragalactic Sources Observed by the FIRST Survey and the Sloan Digital Sky Survey},} \aj, 124, 2364, \dodoi{10.1086/344069}

\bibitem[{{\v{Z}}. {Ivezi{\'c}} {et~al.}(2007){Ivezi{\'c}}, {Smith}, {Miknaitis}, {Lin}, {Tucker}, {Lupton}, {Gunn}, {Knapp}, {Strauss}, {Sesar}, {Doi}, {Tanaka}, {Fukugita}, {Holtzman}, {Kent}, {Yanny}, {Schlegel}, {Finkbeiner}, {Padmanabhan}, {Rockosi}, {Juri{\'c}}, {Bond}, {Lee}, {Stoughton}, {Jester}, {Harris}, {Harding}, {Morrison}, {Brinkmann}, {Schneider}, \& {York}}]{ivezic_2007}
{Ivezi{\'c}}, {\v{Z}}., {Smith}, J.~A., {Miknaitis}, G., {et~al.} 2007, \bibinfo{title}{{Sloan Digital Sky Survey Standard Star Catalog for Stripe 82: The Dawn of Industrial 1\% Optical Photometry},} \aj, 134, 973, \dodoi{10.1086/519976}

\bibitem[{N. {Jiang} {et~al.}(2021){Jiang}, {Wang}, {Hu}, {Sun}, {Dou}, \& {Xiao}}]{jiang_2021}
{Jiang}, N., {Wang}, T., {Hu}, X., {et~al.} 2021, \bibinfo{title}{{Infrared Echoes of Optical Tidal Disruption Events: {\ensuremath{\sim}}1\% Dust-covering Factor or Less at Subparsec Scale},} \apj, 911, 31, \dodoi{10.3847/1538-4357/abe772}

\bibitem[{G. {Kauffmann} {et~al.}(2003{\natexlab{a}}){Kauffmann}, {Heckman}, {Tremonti}, {Brinchmann}, {Charlot}, {White}, {Ridgway}, {Brinkmann}, {Fukugita}, {Hall}, {Ivezi{\'c}}, {Richards}, \& {Schneider}}]{kauffmann_2003}
{Kauffmann}, G., {Heckman}, T.~M., {Tremonti}, C., {et~al.} 2003{\natexlab{a}}, \bibinfo{title}{{The host galaxies of active galactic nuclei},} \mnras, 346, 1055, \dodoi{10.1111/j.1365-2966.2003.07154.x}

\bibitem[{G. {Kauffmann} {et~al.}(2003{\natexlab{b}}){Kauffmann}, {Heckman}, {White}, {Charlot}, {Tremonti}, {Brinchmann}, {Bruzual}, {Peng}, {Seibert}, {Bernardi}, {Blanton}, {Brinkmann}, {Castander}, {Cs{\'a}bai}, {Fukugita}, {Ivezic}, {Munn}, {Nichol}, {Padmanabhan}, {Thakar}, {Weinberg}, \& {York}}]{2003MNRAS.341...33K}
{Kauffmann}, G., {Heckman}, T.~M., {White}, S. D.~M., {et~al.} 2003{\natexlab{b}}, \bibinfo{title}{{Stellar masses and star formation histories for {}10$^{5}$ galaxies from the Sloan Digital Sky Survey},} \mnras, 341, 33, \dodoi{10.1046/j.1365-8711.2003.06291.x}

\bibitem[{K.~I. {Kellermann} {et~al.}(1989){Kellermann}, {Sramek}, {Schmidt}, {Shaffer}, \& {Green}}]{kellermann_1989}
{Kellermann}, K.~I., {Sramek}, R., {Schmidt}, M., {Shaffer}, D.~B., \& {Green}, R. 1989, \bibinfo{title}{{VLA Observations of Objects in the Palomar Bright Quasar Survey},} \aj, 98, 1195, \dodoi{10.1086/115207}

\bibitem[{B.~C. {Kelly} {et~al.}(2009){Kelly}, {Bechtold}, \& {Siemiginowska}}]{kelly_2009}
{Kelly}, B.~C., {Bechtold}, J., \& {Siemiginowska}, A. 2009, \bibinfo{title}{{Are the Variations in Quasar Optical Flux Driven by Thermal Fluctuations?},} \apj, 698, 895, \dodoi{10.1088/0004-637X/698/1/895}

\bibitem[{L.~J. {Kewley} {et~al.}(2001){Kewley}, {Dopita}, {Sutherland}, {Heisler}, \& {Trevena}}]{kewley_2001}
{Kewley}, L.~J., {Dopita}, M.~A., {Sutherland}, R.~S., {Heisler}, C.~A., \& {Trevena}, J. 2001, \bibinfo{title}{{Theoretical Modeling of Starburst Galaxies},} \apj, 556, 121, \dodoi{10.1086/321545}

\bibitem[{M. {Kim} {et~al.}(2021){Kim}, {Barth}, {Ho}, \& {Son}}]{kim_2021}
{Kim}, M., {Barth}, A.~J., {Ho}, L.~C., \& {Son}, S. 2021, \bibinfo{title}{{A Hubble Space Telescope Imaging Survey of Low-redshift Swift-BAT Active Galaxies},} \apjs, 256, 40, \dodoi{10.3847/1538-4365/ac133e}

\bibitem[{M. {Kim} \& L.~C. {Ho}(2019){Kim} \& {Ho}}]{kim_2019}
{Kim}, M., \& {Ho}, L.~C. 2019, \bibinfo{title}{{Evidence for a Young Stellar Population in Nearby Type 1 Active Galaxies},} \apj, 876, 35, \dodoi{10.3847/1538-4357/ab11cf}

\bibitem[{M. {Kim} {et~al.}(2024){Kim}, {Son}, \& {Ho}}]{kim_2024}
{Kim}, M., {Son}, S., \& {Ho}, L.~C. 2024, \bibinfo{title}{{The size-luminosity relation of the AGN torus determined from the comparison between optical and mid-infrared variability},} \aap, 689, A27, \dodoi{10.1051/0004-6361/202450413}

\bibitem[{C.~S. {Kochanek} {et~al.}(2017){Kochanek}, {Shappee}, {Stanek}, {Holoien}, {Thompson}, {Prieto}, {Dong}, {Shields}, {Will}, {Britt}, {Perzanowski}, \& {Pojma{\'n}ski}}]{kochanek_2017}
{Kochanek}, C.~S., {Shappee}, B.~J., {Stanek}, K.~Z., {et~al.} 2017, \bibinfo{title}{{The All-Sky Automated Survey for Supernovae (ASAS-SN) Light Curve Server v1.0},} \pasp, 129, 104502, \dodoi{10.1088/1538-3873/aa80d9}

\bibitem[{M. {Kong} \& L.~C. {Ho}(2018){Kong} \& {Ho}}]{kong_2018}
{Kong}, M., \& {Ho}, L.~C. 2018, \bibinfo{title}{{The Black Hole Masses and Eddington Ratios of Type 2 Quasars},} \apj, 859, 116, \dodoi{10.3847/1538-4357/aabe2a}

\bibitem[{J. {Kormendy} \& L.~C. {Ho}(2013){Kormendy} \& {Ho}}]{kormendy_2013}
{Kormendy}, J., \& {Ho}, L.~C. 2013, \bibinfo{title}{{Coevolution (Or Not) of Supermassive Black Holes and Host Galaxies},} \araa, 51, 511, \dodoi{10.1146/annurev-astro-082708-101811}

\bibitem[{M.~J. {Koss} {et~al.}(2022){Koss}, {Trakhtenbrot}, {Ricci}, {Bauer}, {Treister}, {Mushotzky}, {Urry}, {Ananna}, {Balokovi{\'c}}, {den Brok}, {Cenko}, {Harrison}, {Ichikawa}, {Lamperti}, {Lein}, {Mej{\'\i}a-Restrepo}, {Oh}, {Pacucci}, {Pfeifle}, {Powell}, {Privon}, {Ricci}, {Salvato}, {Schawinski}, {Shimizu}, {Smith}, \& {Stern}}]{koss_2022}
{Koss}, M.~J., {Trakhtenbrot}, B., {Ricci}, C., {et~al.} 2022, \bibinfo{title}{{BASS. XXI. The Data Release 2 Overview},} \apjs, 261, 1, \dodoi{10.3847/1538-4365/ac6c8f}

\bibitem[{S. {Koz{\l}owski} {et~al.}(2010){Koz{\l}owski}, {Kochanek}, {Stern}, {Ashby}, {Assef}, {Bock}, {Borys}, {Brand}, {Brodwin}, {Brown}, {Cool}, {Cooray}, {Croft}, {Dey}, {Eisenhardt}, {Gonzalez}, {Gorjian}, {Griffith}, {Grogin}, {Ivison}, {Jacob}, {Jannuzi}, {Mainzer}, {Moustakas}, {R{\"o}ttgering}, {Seymour}, {Smith}, {Stanford}, {Stauffer}, {Sullivan}, {van Breugel}, {Willner}, \& {Wright}}]{kozlowski_2010}
{Koz{\l}owski}, S., {Kochanek}, C.~S., {Stern}, D., {et~al.} 2010, \bibinfo{title}{{Mid-infrared Variability from the Spitzer Deep Wide-field Survey},} \apj, 716, 530, \dodoi{10.1088/0004-637X/716/1/530}

\bibitem[{M. {Lacy} {et~al.}(2004){Lacy}, {Storrie-Lombardi}, {Sajina}, {Appleton}, {Armus}, {Chapman}, {Choi}, {Fadda}, {Fang}, {Frayer}, {Heinrichsen}, {Helou}, {Im}, {Marleau}, {Masci}, {Shupe}, {Soifer}, {Surace}, {Teplitz}, {Wilson}, \& {Yan}}]{lacy_2004}
{Lacy}, M., {Storrie-Lombardi}, L.~J., {Sajina}, A., {et~al.} 2004, \bibinfo{title}{{Obscured and Unobscured Active Galactic Nuclei in the Spitzer Space Telescope First Look Survey},} \apjs, 154, 166, \dodoi{10.1086/422816}

\bibitem[{M. {Lacy} {et~al.}(2020){Lacy}, {Baum}, {Chandler}, {Chatterjee}, {Clarke}, {Deustua}, {English}, {Farnes}, {Gaensler}, {Gugliucci}, {Hallinan}, {Kent}, {Kimball}, {Law}, {Lazio}, {Marvil}, {Mao}, {Medlin}, {Mooley}, {Murphy}, {Myers}, {Osten}, {Richards}, {Rosolowsky}, {Rudnick}, {Schinzel}, {Sivakoff}, {Sjouwerman}, {Taylor}, {White}, {Wrobel}, {Andernach}, {Beasley}, {Berger}, {Bhatnager}, {Birkinshaw}, {Bower}, {Brandt}, {Brown}, {Burke-Spolaor}, {Butler}, {Comerford}, {Demorest}, {Fu}, {Giacintucci}, {Golap}, {G{\"u}th}, {Hales}, {Hiriart}, {Hodge}, {Horesh}, {Ivezi{\'c}}, {Jarvis}, {Kamble}, {Kassim}, {Liu}, {Loinard}, {Lyons}, {Masters}, {Mezcua}, {Moellenbrock}, {Mroczkowski}, {Nyland}, {O'Dea}, {O'Sullivan}, {Peters}, {Radford}, {Rao}, {Robnett}, {Salcido}, {Shen}, {Sobotka}, {Witz}, {Vaccari}, {van Weeren}, {Vargas}, {Williams}, \& {Yoon}}]{lacy_2020}
{Lacy}, M., {Baum}, S.~A., {Chandler}, C.~J., {et~al.} 2020, \bibinfo{title}{{The Karl G. Jansky Very Large Array Sky Survey (VLASS). Science Case and Survey Design},} \pasp, 132, 035001, \dodoi{10.1088/1538-3873/ab63eb}

\bibitem[{A. {Lamastra} {et~al.}(2009){Lamastra}, {Bianchi}, {Matt}, {Perola}, {Barcons}, \& {Carrera}}]{lamastra_2009}
{Lamastra}, A., {Bianchi}, S., {Matt}, G., {et~al.} 2009, \bibinfo{title}{{The bolometric luminosity of type 2 AGN from extinction-corrected [OIII]. No evidence of Eddington-limited sources},} \aap, 504, 73, \dodoi{10.1051/0004-6361/200912023}

\bibitem[{W. {Li} {et~al.}(2011){Li}, {Leaman}, {Chornock}, {Filippenko}, {Poznanski}, {Ganeshalingam}, {Wang}, {Modjaz}, {Jha}, {Foley}, \& {Smith}}]{li_2011}
{Li}, W., {Leaman}, J., {Chornock}, R., {et~al.} 2011, \bibinfo{title}{{Nearby supernova rates from the Lick Observatory Supernova Search - II. The observed luminosity functions and fractions of supernovae in a complete sample},} \mnras, 412, 1441, \dodoi{10.1111/j.1365-2966.2011.18160.x}

\bibitem[{Y.~A. {Li} {et~al.}(2023){Li}, {Ho}, {Shangguan}, {Zhuang}, \& {Li}}]{li_2023}
{Li}, Y.~A., {Ho}, L.~C., {Shangguan}, J., {Zhuang}, M.-Y., \& {Li}, R. 2023, \bibinfo{title}{{Panchromatic Photometry of Low-redshift, Massive Galaxies Selected from SDSS Stripe 82},} \apjs, 267, 17, \dodoi{10.3847/1538-4365/acd4b5}

\bibitem[{C.~J. {Lonsdale} {et~al.}(2003){Lonsdale}, {Smith}, {Rowan-Robinson}, {Surace}, {Shupe}, {Xu}, {Oliver}, {Padgett}, {Fang}, {Conrow}, {Franceschini}, {Gautier}, {Griffin}, {Hacking}, {Masci}, {Morrison}, {O'Linger}, {Owen}, {P{\'e}rez-Fournon}, {Pierre}, {Puetter}, {Stacey}, {Castro}, {Polletta}, {Farrah}, {Jarrett}, {Frayer}, {Siana}, {Babbedge}, {Dye}, {Fox}, {Gonzalez-Solares}, {Salaman}, {Berta}, {Condon}, {Dole}, \& {Serjeant}}]{lonsdale_2003}
{Lonsdale}, C.~J., {Smith}, H.~E., {Rowan-Robinson}, M., {et~al.} 2003, \bibinfo{title}{{SWIRE: The SIRTF Wide-Area Infrared Extragalactic Survey},} \pasp, 115, 897, \dodoi{10.1086/376850}

\bibitem[{A. {Lupi} {et~al.}(2020){Lupi}, {Sbarrato}, \& {Carniani}}]{lupi_2020}
{Lupi}, A., {Sbarrato}, T., \& {Carniani}, S. 2020, \bibinfo{title}{{Difficulties in mid-infrared selection of AGNs in dwarf galaxies},} \mnras, 492, 2528, \dodoi{10.1093/mnras/stz3636}

\bibitem[{J. {Lyu} {et~al.}(2024){Lyu}, {Alberts}, {Rieke}, {Shivaei}, {P{\'e}rez-Gonz{\'a}lez}, {Sun}, {Hainline}, {Baum}, {Bonaventura}, {Bunker}, {Egami}, {Eisenstein}, {Florian}, {Ji}, {Johnson}, {Morrison}, {Rieke}, {Robertson}, {Rujopakarn}, {Tacchella}, {Scholtz}, \& {Willmer}}]{lyu_2024}
{Lyu}, J., {Alberts}, S., {Rieke}, G.~H., {et~al.} 2024, \bibinfo{title}{{Active Galactic Nuclei Selection and Demographics: A New Age with JWST/MIRI},} \apj, 966, 229, \dodoi{10.3847/1538-4357/ad3643}

\bibitem[{A. {Mainzer} {et~al.}(2011){Mainzer}, {Bauer}, {Grav}, {Masiero}, {Cutri}, {Dailey}, {Eisenhardt}, {McMillan}, {Wright}, {Walker}, {Jedicke}, {Spahr}, {Tholen}, {Alles}, {Beck}, {Brandenburg}, {Conrow}, {Evans}, {Fowler}, {Jarrett}, {Marsh}, {Masci}, {McCallon}, {Wheelock}, {Wittman}, {Wyatt}, {DeBaun}, {Elliott}, {Elsbury}, {Gautier}, {Gomillion}, {Leisawitz}, {Maleszewski}, {Micheli}, \& {Wilkins}}]{mainzer_2011}
{Mainzer}, A., {Bauer}, J., {Grav}, T., {et~al.} 2011, \bibinfo{title}{{Preliminary Results from NEOWISE: An Enhancement to the Wide-field Infrared Survey Explorer for Solar System Science},} \apj, 731, 53, \dodoi{10.1088/0004-637X/731/1/53}

\bibitem[{A. {Mainzer} {et~al.}(2014){Mainzer}, {Bauer}, {Cutri}, {Grav}, {Masiero}, {Beck}, {Clarkson}, {Conrow}, {Dailey}, {Eisenhardt}, {Fabinsky}, {Fajardo-Acosta}, {Fowler}, {Gelino}, {Grillmair}, {Heinrichsen}, {Kendall}, {Kirkpatrick}, {Liu}, {Masci}, {McCallon}, {Nugent}, {Papin}, {Rice}, {Royer}, {Ryan}, {Sevilla}, {Sonnett}, {Stevenson}, {Thompson}, {Wheelock}, {Wiemer}, {Wittman}, {Wright}, \& {Yan}}]{mainzer_2014}
{Mainzer}, A., {Bauer}, J., {Cutri}, R.~M., {et~al.} 2014, \bibinfo{title}{{Initial Performance of the NEOWISE Reactivation Mission},} \apj, 792, 30, \dodoi{10.1088/0004-637X/792/1/30}

\bibitem[{F. {Mannucci} {et~al.}(2005){Mannucci}, {Della Valle}, {Panagia}, {Cappellaro}, {Cresci}, {Maiolino}, {Petrosian}, \& {Turatto}}]{mannucci_2005}
{Mannucci}, F., {Della Valle}, M., {Panagia}, N., {et~al.} 2005, \bibinfo{title}{{The supernova rate per unit mass},} \aap, 433, 807, \dodoi{10.1051/0004-6361:20041411}

\bibitem[{S. {Mateos} {et~al.}(2012){Mateos}, {Alonso-Herrero}, {Carrera}, {Blain}, {Watson}, {Barcons}, {Braito}, {Severgnini}, {Donley}, \& {Stern}}]{mateos_2012}
{Mateos}, S., {Alonso-Herrero}, A., {Carrera}, F.~J., {et~al.} 2012, \bibinfo{title}{{Using the Bright Ultrahard XMM-Newton survey to define an IR selection of luminous AGN based on WISE colours},} \mnras, 426, 3271, \dodoi{10.1111/j.1365-2966.2012.21843.x}

\bibitem[{J.-C. {Mauduit} {et~al.}(2012){Mauduit}, {Lacy}, {Farrah}, {Surace}, {Jarvis}, {Oliver}, {Maraston}, {Vaccari}, {Marchetti}, {Zeimann}, {Gonz{\'a}les-Solares}, {Pforr}, {Petric}, {Henriques}, {Thomas}, {Afonso}, {Rettura}, {Wilson}, {Falder}, {Geach}, {Huynh}, {Norris}, {Seymour}, {Richards}, {Stanford}, {Alexander}, {Becker}, {Best}, {Bizzocchi}, {Bonfield}, {Castro}, {Cava}, {Chapman}, {Christopher}, {Clements}, {Covone}, {Dubois}, {Dunlop}, {Dyke}, {Edge}, {Ferguson}, {Foucaud}, {Franceschini}, {Gal}, {Grant}, {Grossi}, {Hatziminaoglou}, {Hickey}, {Hodge}, {Huang}, {Ivison}, {Kim}, {LeFevre}, {Lehnert}, {Lonsdale}, {Lubin}, {McLure}, {Messias}, {Mart{\'\i}nez-Sansigre}, {Mortier}, {Nielsen}, {Ouchi}, {Parish}, {Perez-Fournon}, {Pierre}, {Rawlings}, {Readhead}, {Ridgway}, {Rigopoulou}, {Romer}, {Rosebloom}, {Rottgering}, {Rowan-Robinson}, {Sajina}, {Simpson}, {Smail}, {Squires}, {Stevens}, {Taylor}, {Trichas}, {Urrutia}, {van Kampen}, {Verma}, \& {Xu}}]{mauduit_2012}
{Mauduit}, J.-C., {Lacy}, M., {Farrah}, D., {et~al.} 2012, \bibinfo{title}{{The Spitzer Extragalactic Representative Volume Survey (SERVS): Survey Definition and Goals},} \pasp, 124, 714, \dodoi{10.1086/666945}

\bibitem[{M.~A. {McLaughlin} {et~al.}(1996){McLaughlin}, {Mattox}, {Cordes}, \& {Thompson}}]{mclaughlin_1996}
{McLaughlin}, M.~A., {Mattox}, J.~R., {Cordes}, J.~M., \& {Thompson}, D.~J. 1996, \bibinfo{title}{{Variability of CGRO/EGRET Gamma-Ray Sources},} \apj, 473, 763, \dodoi{10.1086/178188}

\bibitem[{R. {Mor} \& B. {Trakhtenbrot}(2011){Mor} \& {Trakhtenbrot}}]{mor_2011}
{Mor}, R., \& {Trakhtenbrot}, B. 2011, \bibinfo{title}{{Hot-dust Clouds with Pure-graphite Composition around type-I Active Galactic Nuclei},} \apjl, 737, L36, \dodoi{10.1088/2041-8205/737/2/L36}

\bibitem[{K.~G. {Noeske} {et~al.}(2007){Noeske}, {Weiner}, {Faber}, {Papovich}, {Koo}, {Somerville}, {Bundy}, {Conselice}, {Newman}, {Schiminovich}, {Le Floc'h}, {Coil}, {Rieke}, {Lotz}, {Primack}, {Barmby}, {Cooper}, {Davis}, {Ellis}, {Fazio}, {Guhathakurta}, {Huang}, {Kassin}, {Martin}, {Phillips}, {Rich}, {Small}, {Willmer}, \& {Wilson}}]{noeske_2007}
{Noeske}, K.~G., {Weiner}, B.~J., {Faber}, S.~M., {et~al.} 2007, \bibinfo{title}{{Star Formation in AEGIS Field Galaxies since z=1.1: The Dominance of Gradually Declining Star Formation, and the Main Sequence of Star-forming Galaxies},} \apjl, 660, L43, \dodoi{10.1086/517926}

\bibitem[{K. {Oh} {et~al.}(2018){Oh}, {Koss}, {Markwardt}, {Schawinski}, {Baumgartner}, {Barthelmy}, {Cenko}, {Gehrels}, {Mushotzky}, {Petulante}, {Ricci}, {Lien}, \& {Trakhtenbrot}}]{oh_2018}
{Oh}, K., {Koss}, M., {Markwardt}, C.~B., {et~al.} 2018, \bibinfo{title}{{The 105-Month Swift-BAT All-sky Hard X-Ray Survey},} \apjs, 235, 4, \dodoi{10.3847/1538-4365/aaa7fd}

\bibitem[{A. {Pai} {et~al.}(2024){Pai}, {Blanton}, \& {Moustakas}}]{pai_2024}
{Pai}, A., {Blanton}, M.~R., \& {Moustakas}, J. 2024, \bibinfo{title}{{Mid-infrared Variability in Nearby Galaxies from the MaNGA Sample},} \apj, 977, 102, \dodoi{10.3847/1538-4357/ad89b8}

\bibitem[{T. pandas~development team(2020)pandas~development team}]{2020pandas}
pandas~development team, T. 2020, pandas-dev/pandas: Pandas, latest Zenodo, \dodoi{10.5281/zenodo.3509134}

\bibitem[{F. Pedregosa {et~al.}(2011)Pedregosa, Varoquaux, Gramfort, Michel, Thirion, Grisel, Blondel, Prettenhofer, Weiss, Dubourg, Vanderplas, Passos, Cournapeau, Brucher, Perrot, \& Duchesnay}]{2011scikit-learn}
Pedregosa, F., Varoquaux, G., Gramfort, A., {et~al.} 2011, \bibinfo{title}{Scikit-learn: Machine Learning in {P}ython,} Journal of Machine Learning Research, 12, 2825

\bibitem[{A. {Renzini} \& Y.-j. {Peng}(2015){Renzini} \& {Peng}}]{renzini_2015}
{Renzini}, A., \& {Peng}, Y.-j. 2015, \bibinfo{title}{{An Objective Definition for the Main Sequence of Star-forming Galaxies},} \apjl, 801, L29, \dodoi{10.1088/2041-8205/801/2/L29}

\bibitem[{G.~T. {Richards} {et~al.}(2002){Richards}, {Fan}, {Newberg}, {Strauss}, {Vanden Berk}, {Schneider}, {Yanny}, {Boucher}, {Burles}, {Frieman}, {Gunn}, {Hall}, {Ivezi{\'c}}, {Kent}, {Loveday}, {Lupton}, {Rockosi}, {Schlegel}, {Stoughton}, {SubbaRao}, \& {York}}]{richards_2002}
{Richards}, G.~T., {Fan}, X., {Newberg}, H.~J., {et~al.} 2002, \bibinfo{title}{{Spectroscopic Target Selection in the Sloan Digital Sky Survey: The Quasar Sample},} \aj, 123, 2945, \dodoi{10.1086/340187}

\bibitem[{J. {Sabater} {et~al.}(2019){Sabater}, {Best}, {Hardcastle}, {Shimwell}, {Tasse}, {Williams}, {Br{\"u}ggen}, {Cochrane}, {Croston}, {de Gasperin}, {Duncan}, {G{\"u}rkan}, {Mechev}, {Morabito}, {Prandoni}, {R{\"o}ttgering}, {Smith}, {Harwood}, {Mingo}, {Mooney}, \& {Saxena}}]{sabater_2019}
{Sabater}, J., {Best}, P.~N., {Hardcastle}, M.~J., {et~al.} 2019, \bibinfo{title}{{The LoTSS view of radio AGN in the local Universe. The most massive galaxies are always switched on},} \aap, 622, A17, \dodoi{10.1051/0004-6361/201833883}

\bibitem[{A. {Saintonge} {et~al.}(2016){Saintonge}, {Catinella}, {Cortese}, {Genzel}, {Giovanelli}, {Haynes}, {Janowiecki}, {Kramer}, {Lutz}, {Schiminovich}, {Tacconi}, {Wuyts}, \& {Accurso}}]{saintonge_2016}
{Saintonge}, A., {Catinella}, B., {Cortese}, L., {et~al.} 2016, \bibinfo{title}{{Molecular and atomic gas along and across the main sequence of star-forming galaxies},} \mnras, 462, 1749, \dodoi{10.1093/mnras/stw1715}

\bibitem[{S. {Salim} {et~al.}(2007){Salim}, {Rich}, {Charlot}, {Brinchmann}, {Johnson}, {Schiminovich}, {Seibert}, {Mallery}, {Heckman}, {Forster}, {Friedman}, {Martin}, {Morrissey}, {Neff}, {Small}, {Wyder}, {Bianchi}, {Donas}, {Lee}, {Madore}, {Milliard}, {Szalay}, {Welsh}, \& {Yi}}]{salim_2007}
{Salim}, S., {Rich}, R.~M., {Charlot}, S., {et~al.} 2007, \bibinfo{title}{{UV Star Formation Rates in the Local Universe},} \apjs, 173, 267, \dodoi{10.1086/519218}

\bibitem[{P. {S{\'a}nchez} {et~al.}(2017){S{\'a}nchez}, {Lira}, {Cartier}, {P{\'e}rez}, {Miranda}, {Yovaniniz}, {Ar{\'e}valo}, {Milvang-Jensen}, {Fynbo}, {Dunlop}, {Coppi}, \& {Marchesi}}]{sanchez_2017}
{S{\'a}nchez}, P., {Lira}, P., {Cartier}, R., {et~al.} 2017, \bibinfo{title}{{Near-infrared Variability of Obscured and Unobscured X-Ray-selected AGNs in the COSMOS Field},} \apj, 849, 110, \dodoi{10.3847/1538-4357/aa9188}

\bibitem[{R. {She} {et~al.}(2017){She}, {Ho}, \& {Feng}}]{she_2017}
{She}, R., {Ho}, L.~C., \& {Feng}, H. 2017, \bibinfo{title}{{Chandra Survey of Nearby Galaxies: The Catalog},} \apj, 835, 223, \dodoi{10.3847/1538-4357/835/2/223}

\bibitem[{J. {Silk} \& M.~J. {Rees}(1998){Silk} \& {Rees}}]{silk_1998}
{Silk}, J., \& {Rees}, M.~J. 1998, \bibinfo{title}{{Quasars and galaxy formation},} \aap, 331, L1, \dodoi{10.48550/arXiv.astro-ph/9801013}

\bibitem[{M. {Siudek} {et~al.}(2025){Siudek}, {Mezcua}, {Circosta}, {Maraston}, {Moustakas}, {Zou}, {Aguilar}, {Ahlen}, {Bianchi}, {Brooks}, {Claybaugh}, {Dawson}, {de la Macorra}, {Dey}, {Doel}, {Forero-Romero}, {Gazta{\~n}aga}, {Gontcho A Gontcho}, {Gutierrez}, {Ishak}, {Juneau}, {Kirkby}, {Kisner}, {Kremin}, {Lambert}, {Landriau}, {Le Guillou}, {Meisner}, {Miquel}, {Prada}, {P{\'e}rez-R{\`a}fols}, {Rossi}, {Sanchez}, {Schlegel}, {Schubnell}, {Seo}, {Sprayberry}, {Tarl{\'e}}, \& {Weaver}}]{siudek_2025}
{Siudek}, M., {Mezcua}, M., {Circosta}, C., {et~al.} 2025, \bibinfo{title}{{Beyond traditional diagnostics: Identifying active galactic nuclei using spectral energy distribution fitting in DESI data},} \aap, 700, A209, \dodoi{10.1051/0004-6361/202555463}

\bibitem[{S. {Son} {et~al.}(2022){Son}, {Kim}, \& {Ho}}]{son_2022}
{Son}, S., {Kim}, M., \& {Ho}, L.~C. 2022, \bibinfo{title}{{Mid-infrared Variability of Low-redshift Active Galactic Nuclei: Constraints on a Hot Dust Component with a Variable Covering Factor},} \apj, 927, 107, \dodoi{10.3847/1538-4357/ac4dfc}

\bibitem[{S. {Son} {et~al.}(2023){Son}, {Kim}, \& {Ho}}]{son_2023}
{Son}, S., {Kim}, M., \& {Ho}, L.~C. 2023, \bibinfo{title}{{The Structure Function of Mid-infrared Variability in Low-redshift Active Galactic Nuclei},} \apj, 958, 135, \dodoi{10.3847/1538-4357/ad01bc}

\bibitem[{S. {Son} {et~al.}(2025){Son}, {Kim}, \& {Ho}}]{son_2025}
{Son}, S., {Kim}, M., \& {Ho}, L.~C. 2025, \bibinfo{title}{{Temperature profiles of accretion disks in luminous active galactic nuclei derived from ultraviolet spectroscopic variability},} \aap, 695, A268, \dodoi{10.1051/0004-6361/202452467}

\bibitem[{S. {Son} {et~al.}(2026){Son}, {Kim}, \& {Ho}}]{son_2026}
{Son}, S., {Kim}, M., \& {Ho}, L.~C. 2026, \bibinfo{title}{{Asymmetric torus variability in active galactic nuclei driven by global brightening and dimming},} \aap, 706, A122, \dodoi{10.1051/0004-6361/202557499}

\bibitem[{D. {Stern} {et~al.}(2005){Stern}, {Eisenhardt}, {Gorjian}, {Kochanek}, {Caldwell}, {Eisenstein}, {Brodwin}, {Brown}, {Cool}, {Dey}, {Green}, {Jannuzi}, {Murray}, {Pahre}, \& {Willner}}]{stern_2005}
{Stern}, D., {Eisenhardt}, P., {Gorjian}, V., {et~al.} 2005, \bibinfo{title}{{Mid-Infrared Selection of Active Galaxies},} \apj, 631, 163, \dodoi{10.1086/432523}

\bibitem[{D. {Stern} {et~al.}(2012){Stern}, {Assef}, {Benford}, {Blain}, {Cutri}, {Dey}, {Eisenhardt}, {Griffith}, {Jarrett}, {Lake}, {Masci}, {Petty}, {Stanford}, {Tsai}, {Wright}, {Yan}, {Harrison}, \& {Madsen}}]{stern_2012}
{Stern}, D., {Assef}, R.~J., {Benford}, D.~J., {et~al.} 2012, \bibinfo{title}{{Mid-infrared Selection of Active Galactic Nuclei with the Wide-Field Infrared Survey Explorer. I. Characterizing WISE-selected Active Galactic Nuclei in COSMOS},} \apj, 753, 30, \dodoi{10.1088/0004-637X/753/1/30}

\bibitem[{N.~C. {Stone} \& B.~D. {Metzger}(2016){Stone} \& {Metzger}}]{stone_2016}
{Stone}, N.~C., \& {Metzger}, B.~D. 2016, \bibinfo{title}{{Rates of stellar tidal disruption as probes of the supermassive black hole mass function},} \mnras, 455, 859, \dodoi{10.1093/mnras/stv2281}

\bibitem[{T. {Szalai} {et~al.}(2019){Szalai}, {Zs{\'\i}ros}, {Fox}, {Pejcha}, \& {M{\"u}ller}}]{szalai_2019}
{Szalai}, T., {Zs{\'\i}ros}, S., {Fox}, O.~D., {Pejcha}, O., \& {M{\"u}ller}, T. 2019, \bibinfo{title}{{A Comprehensive Analysis of Spitzer Supernovae},} \apjs, 241, 38, \dodoi{10.3847/1538-4365/ab10df}

\bibitem[{E. {Treister} \& C.~M. {Urry}(2006){Treister} \& {Urry}}]{treister_2006}
{Treister}, E., \& {Urry}, C.~M. 2006, \bibinfo{title}{{The Evolution of Obscuration in Active Galactic Nuclei},} \apjl, 652, L79, \dodoi{10.1086/510237}

\bibitem[{S. {Veilleux} \& D.~E. {Osterbrock}(1987){Veilleux} \& {Osterbrock}}]{veilleux_1987}
{Veilleux}, S., \& {Osterbrock}, D.~E. 1987, \bibinfo{title}{{Spectral Classification of Emission-Line Galaxies},} \apjs, 63, 295, \dodoi{10.1086/191166}

\bibitem[{M. {Venanzi} {et~al.}(2020){Venanzi}, {H{\"o}nig}, \& {Williamson}}]{venanzi_2020}
{Venanzi}, M., {H{\"o}nig}, S., \& {Williamson}, D. 2020, \bibinfo{title}{{The Role of Infrared Radiation Pressure in Shaping Dusty Winds in AGNs},} \apj, 900, 174, \dodoi{10.3847/1538-4357/aba89f}

\bibitem[{P. Virtanen {et~al.}(2020)Virtanen, Gommers, Oliphant, Haberland, Reddy, Cournapeau, Burovski, Peterson, Weckesser, Bright, {van der Walt}, Brett, Wilson, Millman, Mayorov, Nelson, Jones, Kern, Larson, Carey, Polat, Feng, Moore, {VanderPlas}, Laxalde, Perktold, Cimrman, Henriksen, Quintero, Harris, Archibald, Ribeiro, Pedregosa, {van Mulbregt}, \& {SciPy 1.0 Contributors}}]{2020SciPy}
Virtanen, P., Gommers, R., Oliphant, T.~E., {et~al.} 2020, \bibinfo{title}{{{SciPy} 1.0: Fundamental Algorithms for Scientific Computing in Python},} Nature Methods, 17, 261, \dodoi{10.1038/s41592-019-0686-2}

\bibitem[{K. {Wada}(2012){Wada}}]{wada_2012}
{Wada}, K. 2012, \bibinfo{title}{{Radiation-driven Fountain and Origin of Torus around Active Galactic Nuclei},} \apj, 758, 66, \dodoi{10.1088/0004-637X/758/1/66}

\bibitem[{ {WISE Team}(2020{\natexlab{a}}){WISE Team}}]{AllWISE}
{WISE Team}. 2020{\natexlab{a}}, AllWISE Multiepoch Photometry Table, IPAC, \dodoi{10.26131/IRSA134}

\bibitem[{ {WISE Team}(2020{\natexlab{b}}){WISE Team}}]{neowise}
{WISE Team}. 2020{\natexlab{b}}, NEOWISE 2-Band Post-Cryo Single Exposure (L1b) Source Table, IPAC, \dodoi{10.26131/IRSA124}

\bibitem[{E.~L. {Wright} {et~al.}(2010){Wright}, {Eisenhardt}, {Mainzer}, {Ressler}, {Cutri}, {Jarrett}, {Kirkpatrick}, {Padgett}, {McMillan}, {Skrutskie}, {Stanford}, {Cohen}, {Walker}, {Mather}, {Leisawitz}, {Gautier}, {McLean}, {Benford}, {Lonsdale}, {Blain}, {Mendez}, {Irace}, {Duval}, {Liu}, {Royer}, {Heinrichsen}, {Howard}, {Shannon}, {Kendall}, {Walsh}, {Larsen}, {Cardon}, {Schick}, {Schwalm}, {Abid}, {Fabinsky}, {Naes}, \& {Tsai}}]{wright_2010}
{Wright}, E.~L., {Eisenhardt}, P. R.~M., {Mainzer}, A.~K., {et~al.} 2010, \bibinfo{title}{{The Wide-field Infrared Survey Explorer (WISE): Mission Description and Initial On-orbit Performance},} \aj, 140, 1868, \dodoi{10.1088/0004-6256/140/6/1868}

\bibitem[{H. {Yuk} {et~al.}(2022){Yuk}, {Dai}, {Jayasinghe}, {Fu}, {Mishra}, {Kochanek}, {Shappee}, \& {Stanek}}]{yuk_2022}
{Yuk}, H., {Dai}, X., {Jayasinghe}, T., {et~al.} 2022, \bibinfo{title}{{Variability Selected Active Galactic Nuclei from ASAS-SN Survey: Constraining the Low Luminosity AGN Population},} \apj, 930, 110, \dodoi{10.3847/1538-4357/ac6423}

\bibitem[{M.-Y. {Zhuang} \& L.~C. {Ho}(2023){Zhuang} \& {Ho}}]{zhuang_2023}
{Zhuang}, M.-Y., \& {Ho}, L.~C. 2023, \bibinfo{title}{{Evolutionary paths of active galactic nuclei and their host galaxies},} Nature Astronomy, 7, 1376, \dodoi{10.1038/s41550-023-02051-4}

\end{thebibliography}
\bibliographystyle{aasjournalv7}

\appendix
\section{Example Light Curves of `Suspicious' Source}

Figure~A1 presents example light curves for the suspicious sources with $P_{\rm var}$ values ranging from 0.95 to 0.99.
\restartappendixnumbering

\begin{figure*}[htp]
\centering
\includegraphics[width=0.97\textwidth]{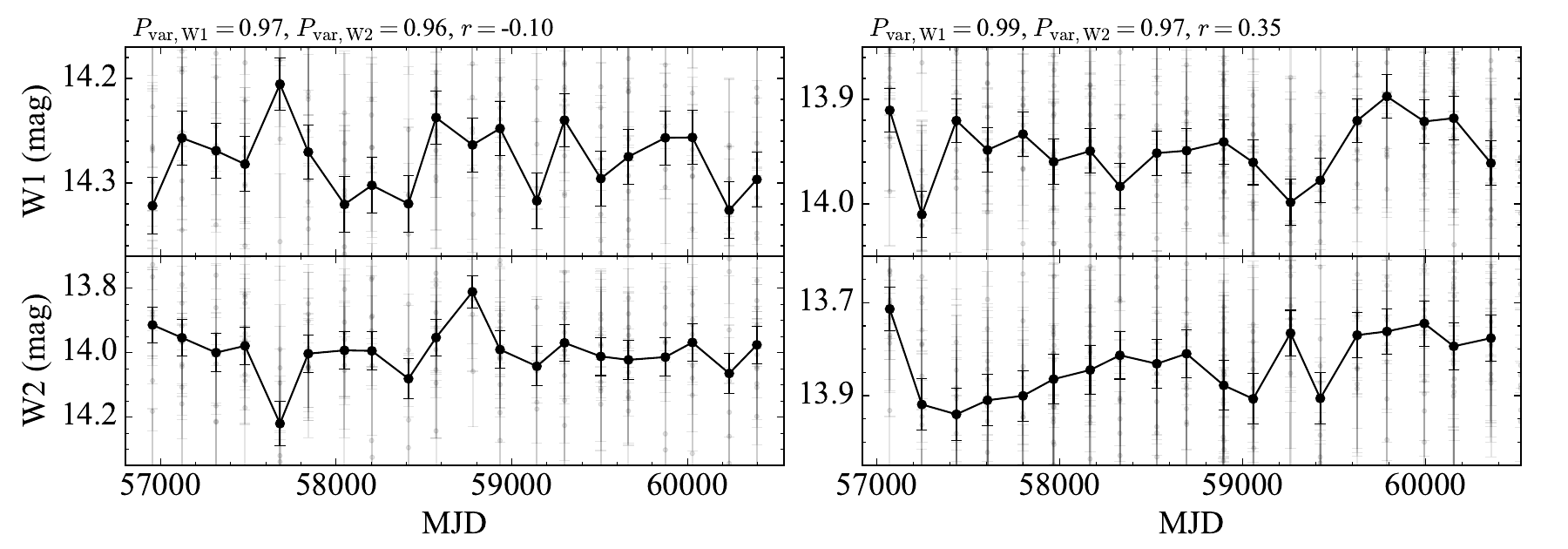}
\caption{ Same as Figure 1, but for suspicious sources with $0.95 < P_{\rm var} < 0.99$.}
\end{figure*}

\end{document}